\documentclass[acmsmall]{acmart}

\usepackage{booktabs}   
\usepackage{subcaption} 

\usepackage{mathtools}

\usepackage{amsmath,array}

\usepackage{amssymb}
\usepackage{bm}

\makeatletter\if@ACM@journal\makeatother
\acmJournal{PACMPL}
\acmVolume{0}
\acmNumber{0}
\acmArticle{0}
\acmYear{2017}
\acmMonth{10}
\acmDOI{Unpublished}
\acmPrice{X}
\startPage{1}
\else\makeatother
\acmConference[]{}{}{}
\acmYear{}
\acmISBN{}
\acmDOI{}
\startPage{1}
\fi

\setcopyright{none}             

\newcommand{\papertitle}{Strategy Preserving Compilation for Parallel Functional Code}

\usepackage{xspace}
\usepackage{stmaryrd}
\usepackage{cmll}
\usepackage{cleveref}
\usepackage{mathpartir}
\usepackage{listings}
\lstdefinelanguage[OpenCL]{C}[ANSI]{C}
{morekeywords={__kernel,kernel,__local,local,__global,global,%
__constant,constant,__private,private,%
char2,char3,char4,char8,char16,%
uchar2,uchar3,uchar4,uchar8,uchar16,%
short2,short3,short4,short8,short16,%
ushort2,ushort3,ushort4,ushort8,ushort16,%
int2,int3,int4,int8,int16,%
uint2,uint3,uint4,uint8,uint16,%
long2,long3,long4,long8,long16,%
ulong2,ulong3,ulong4,ulong8,ulong16,%
float2,float3,float4,float8,float16,%
image2d_t,image3d_t,sampler_t,event_t,%
bool2,bool3,bool4,bool8,bool16,%
half2,half3,half4,half8,half16,%
quad,quad2,quad3,quad4,quad8,quad16,%
complex,imaginary}
}

\lstdefinelanguage{Lift}
{morekeywords={split,join,reduce,map,zip,mapSeq,mapLocal,mapGlb,mapWorkgroup,toLocal,toGlobal,toPrivate,iterate,reduceSeq,gather}%
}

\lstdefinelanguage[pseudo]{C}[ANSI]{C}
{morekeywords={parfor}
}

\newcommand{\phrasetype}{\mathsf{phrase}}
\newcommand{\passivetype}{\mathsf{passive}}
\newcommand{\datatype}{\mathsf{data}}
\newcommand{\natkind}{\mathsf{nat}}
\newcommand{\sem}[1]{\llbracket #1 \rrbracket}

\newcommand{\tycomm}{\mathsf{comm}}
\newcommand{\tyexp}{\mathsf{exp}}
\newcommand{\tyacc}{\mathsf{acc}}
\newcommand{\tyvar}{\mathsf{var}}

\newcommand{\pureto}{\to_{\textrm{p}}}

\newcommand{\DPIA}{\ensuremath{\mathsf{DPIA}}\xspace}

\newcommand{\typ}[5]{#1 \mid #2; #3 \vdash #4 : #5}
\newcommand{\typd}{\typ{\Delta}}

\newcommand{\semA}[1]{\mathcal{A}\llparenthesis #1 \rrparenthesis}
\newcommand{\semE}[1]{\mathcal{C}\llparenthesis #1 \rrparenthesis}

\newcommand{\projone}[1]{#1.1}
\newcommand{\projtwo}[1]{#1.2}

\newcommand{\prim}[1]{\bm{\mathsf{#1}}}

\newcommand{\mapI}{\ensuremath{\prim{mapI}}\xspace}
\newcommand{\reduceI}{\ensuremath{\prim{reduceI}}\xspace}

\newcommand{\eg}{\emph{e.g.},\xspace}

\newcommand{\mathrightfill}{\hskip \textwidth minus \textwidth \strut}

\definecolor{shade}{RGB}{223,223,223}
\newcommand\shade[1]{\setlength{\fboxsep}{0pt}\colorbox{shade}{\ensuremath{#1}}}

\renewenvironment{displaymath}{\small\[}{\]\normalsize\ignorespacesafterend}
\newenvironment{smallequation}{\small\begin{equation}}{\end{equation}\normalsize\ignorespacesafterend}

\newcommand\citeapos[1]{\citeauthor{#1}'s \citeyear{#1}}

\begin{document}

\title{\papertitle}


\author{Robert Atkey}
\orcid{0000-0002-4414-5047}             
\affiliation{
  \position{Lecturer}
  \department{Computer and Information Sciences}              
  \institution{University of Strathclyde}            
  \streetaddress{26 Richmond Street}
  \city{Glasgow}
  \state{Scotland}
  \postcode{G1 1XH}
  \country{United Kingdom}
}
\email{robert.atkey@strath.ac.uk}          

\author{Michel Steuwer}
\orcid{0000-0001-5048-0741}             
\affiliation{
  \position{Research Associate}
  \department{School of Informatics}             
  \institution{University of Edinburgh}           
  \streetaddress{10 Crichton Street}
  \city{Edinburgh}
  \state{Scotland}
  \postcode{EH8 9AB}
  \country{United Kingdom}
}
\email{michel.steuwer@ed.ac.uk}         

\author{Sam Lindley}
\orcid{nnnn-nnnn-nnnn-nnnn}             
\affiliation{
  \position{Research Fellow}
  \department{School of Informatics}             
  \institution{University of Edinburgh}           
  \streetaddress{10 Crichton Street}
  \city{Edinburgh}
  \state{Scotland}
  \postcode{EH8 9AB}
  \country{United Kingdom}
}
\email{sam.lindley@ed.ac.uk}         

\author{Christophe Dubach}
\orcid{nnnn-nnnn-nnnn-nnnn}             
\affiliation{
  \position{Lecturer}
  \department{School of Informatics}             
  \institution{University of Edinburgh}           
  \streetaddress{10 Crichton Street}
  \city{Edinburgh}
  \state{Scotland}
  \postcode{EH8 9AB}
  \country{United Kingdom}
}
\email{michel.steuwer@ed.ac.uk}         


\begin{abstract}

Graphics Processing Units (GPUs) and other parallel devices are widely available and have the potential for accelerating a wide class of algorithms.
However, expert programming skills are required to achieving maximum performance.
These devices expose low-level hardware details through imperative programming interfaces where programmers explicity encode device-specific optimisation strategies.
This inevitably results in non-performance-portable programs delivering suboptimal performance on other devices.

Functional programming models have recently seen a renaissance in the systems community as they offer solutions for tackling the performance portability challenge.
Recent work has shown how to automatically choose high-performance parallelisation strategies for a wide range of hardware architectures encoded in a functional representation.
However, the translation of such functional representations to the imperative program expected by the hardware interface is typically performed ad hoc with no correctness guarantees and no guarantees to preserve the intended parallelisation strategy.

In this paper, we present a formalised \emph{strategy-preserving} translation from high-level functional code to low-level data race free parallel imperative code.
This translation is formulated and proved correct within a language we call Data Parallel Idealised Algol (\DPIA), a dialect of Reynolds' Idealised Algol.
Performance results on GPUs and a multicore CPU show that the formalised translation process generates low-level code with performance on a par with code generated from ad hoc approaches.



\end{abstract}



\maketitle

\section{Introduction}
\label{sec:intro}

Modern parallel hardware such as Graphics Processing Units (GPUs) are difficult to program and optimise for.
They require the use of imperative programming interfaces such as OpenMP, OpenCL, or CUDA, which expose low-level hardware details.
Although these interfaces enable experts to squeeze the last bit of performance from the device, they keep most non-expert programmers at bay.
Furthermore, software must be rewritten and tuned specifically for each new generation of device, leading to performance portability problems.

Recent years have seen an accelerating trend towards high level \emph{functional} programming models for expressing parallel computation.
The absence of side-effects, the use of higher order functions, and the compositionality of functional languages makes them particularly attractive for expressing parallel operations using primitives such as $\prim{map}$ and $\prim{reduce}$.
\textsc{Nesl}~\citep{blelloch93nessl}, CopperHead~\citep{catanzaro11copperhead}, LiquidMetal~\citep{dubach12lime}, Accelerate~\citep{mcdonell13optimising}, and Delite~\citep{sujeeth14delite} are examples of such functional approaches.

While these approaches offer a high level programming abstraction, they struggle to deliver high performance across the wide range of currently available parallel hardware or even across different GPUs.
Ultimately, functional abstractions must be translated into imperative code, as dictated by contemporary parallel hardware interfaces.
However, there is a distinct mismatch between high-level functional abstractions and low-level hardware primitives.
Crutially, the translation has to decide for a parallelisation and optimisation strategy before translating the functional into imperative code which implements this strategy.
Current approaches rely on ad hoc techniques typically using a fixed parallelisation strategy manually optimised for a small set of specific devices.
These approaches often fail to deliver high performance when targeting different types of device due to the limited ability to explore other optimisation strategies.

\citet{SteuwerFeLiDu2015/icfp} showed that it is possible to choose parallelisation and optimisation strategies automatically and represent them using functional abstractions.
Their approach uses a rewrite system that encodes strategic decissions as semantics preserving transformations at the functional level.
The rewrite rules are classified into hardware agnostic algorithmic rules and lowering rules that transform high level primitives into hardware specific functional primitives.
An automated search chooses the parallelisation and optimisation strategy by discovering a sequence of rewrites that yields high performance code.
However, the last step of their process, \emph{the conversion of functional into imperative code}, is still performed in an ad hoc manner with no correctness guarantees. 

In this paper, we formalise an \emph{semantics and strategy preserving translation} from functional to parallel imperative code using \emph{Data Parallel Idealised Algol} (\DPIA), a novel variant of Reynolds' \emph{Idealised Algol} \cite{Reynolds97}.
Idealised Algol orthogonally combines typed functional programming with imperative programming, a property that \DPIA inherits.
This allows us to start in the purely functional subset with a high level specification of the computation we wish to perform, and then to systematically rewrite it to ``purely imperative'' code (\autoref{sec:translation-i}) that has a straightforward translation to the C-like OpenCL code required by GPUs (\autoref{sec:translation-iii}).
\DPIA incorporates a substructural type system, incorporating ideas from Reynolds' \emph{Syntactic Control of Inteference} \cite{Reynolds78,OHearnPTT99}, which ensures that the generated programs are data race free.
We show that our translation from functional to imperative is correct (\autoref{sec:correctness}).
Our experimental results in \autoref{sec:experimentalResults} show that our approach delivers high quality code with performance on a par with hand-tuned OpenCL code on several GPUs, along with a correctness guarantee.

Data Parallel Idealised Algol provides a greater benefit than just a correctness proof. Reynolds' Syntactic Control of Interference discipline provides a framework in which to explore the boundaries between functional and imperative parallel programming. We demonstrate the flexibility of this approach in \autoref{sec:opencl} where we augment the basic design of \DPIA with OpenCL specific primitives that describe how we use the parallelism and memory hierarchies of OpenCL, conveying this usage information through the translation to imperative code, whilst retaining the correctness and data race freedom properties.

\paragraph{Contributions}
This paper makes three major contributions:
\begin{enumerate}
\item We describe a formal translation from high level functional code annotated with parallelism strategies to parallel imperative code. We describe this translation fully in \autoref{sec:translation} and prove it correct in \autoref{sec:correctness}.
\item In order to formulate this translation, we introduce a variant of Reynolds' Idealised Algol, called \emph{Data Parallel Idealised Algol} (\DPIA). This is an extension of \citeapos{OHearnPTT99} \emph{Syntactic Control of Inteference Revisited} with indexed types, array and tuple data types, and primitives for data parallel programming. These extensions are essential for data parallel imperative programs. We describe \DPIA in \autoref{sec:typeSystem}.
\item We specialise our framework to OpenCL, demonstrating how to incorporate the parallelism and memory hierarchy requirements of OpenCL into our methodology (\autoref{sec:opencl}). Our experimental results in \autoref{sec:experimentalResults} demonstrate that our formalised translation yields efficient parallel OpenCL code with no overhead.
\end{enumerate}
In the next section we introduce and motivate our approach and the use of Idealised Algol as a foundation. After the technical body of the paper in Sections \ref{sec:typeSystem}-\ref{sec:experimentalResults}, we discuss related work in \autoref{sec:relatedWork} and our conclusions and future work in \autoref{sec:conclusion}.



\section{Strategy Preserving Compilation}
\label{sec:motivation}

We describe a compilation method that takes high level functional array code to low level parallel imperative code. Our approach is characterised by (a) expression of parallelisation strategies in high level functional code, where their semantic equivalence to the original code can be easily checked, and new strategies readily derived; and (b) a predictable and principled translation to low level imperative code that preserves parallelisation strategies.

\subsection{Expressing Parallelisation Strategies in Functional Code}

Here is an expression that describes the dot product of two vectors $\mathit{xs}$ and $\mathit{ys}$:
\begin{smallequation}
  \label{code-ex:simple-dot-product}
  \prim{reduce}~(+)~0~(\prim{map}~(\lambda x.~\prim{fst}~x*\prim{snd}~x)~(\prim{zip}~\mathit{xs}~\mathit{ys}))
\end{smallequation}
This expression can be read in two ways. Firstly, read mathematically, it is a declarative specification of the dot product. Secondly, it can be read as a strategy for computing dot products. Reading right-to-left, we have a pipeline arrangement. Let us make the following assumptions: \emph{i)}~$\prim{zip}$ is not materialised (it only affects how later parts of the pipeline read their input); \emph{ii)}~$\prim{map}$ is executed in parallel across the array; and \emph{iii)}~$\prim{reduce}$ is executed sequentially. Then we can read this expression as embodying a naive ``parallel map, sequential reduce'' strategy.

Such a naive strategy is not always best.
If we try to execute one parallel job per element of the input arrays, then depending on the underlying architecture we will either fail (e.g., on GPUs with a fixed number of execution units), or generate so many threads that coordination of them will dominate the runtime (e.g., on CPUs).
The overall strategy of ``parallel, then sequential'' is likely not the most efficient, either.

We can give a more refined strategy given information about the underlying architecture.
For instance, GPUs support nesting of parallelism by organising threads into groups, or \emph{work-items} into \emph{work-groups}, using OpenCL terminology.
If we know that the input is of size $n \times 128 \times 2048$, we can explicitly control how parallelism can be mapped to the GPU hierarchy.
The following expression distributes the work among $n$ groups of $128$ local threads, each processing $2048$ elements in one go, by directly reducing over the multiplied pairs of elements:
\begin{smallequation}
  \label{code-ex:dot-product-complex}
  \begin{array}[t]{@{}l}
  \prim{reduce}~(+)~0~
  (\prim{join}~(\prim{mapWorkgroup}\\
  \qquad\qquad\qquad\quad\begin{array}[t]{@{}l}
    (\lambda \mathit{zs}_1.~\prim{mapLocal}~(\lambda\mathit{zs}_2.~\prim{reduce}~
      (\lambda x~a.~(\prim{fst}~x * \prim{snd}~x) + a)~0~(\prim{split}~2048~\mathit{zs}_2))~\mathit{zs}_1) \\
     (\prim{split}~(2048 * 128)~(\prim{zip}~\mathit{xs}~\mathit{ys})))))
  \end{array}
  \end{array}
\end{smallequation}
Although this expression gives much more information about how to process the computation on the GPU, we have not left the functional paradigm, so we still have access to the straightforward mathematical reading of this expression.
We can use equational reasoning to prove that this is semantically equivalent to (\ref{code-ex:simple-dot-product}).
Equational reasoning can also be used to
generate (\ref{code-ex:dot-product-complex}) from (\ref{code-ex:simple-dot-product}).
Indeed \citet{SteuwerFeLiDu2015/icfp} have shown that stochastic search techniques are effective at automatically discovering parallelisation strategies that match hand-coded ones.

However, even with a specified parallelisation strategy we cannot execute this code directly.
We need to translate the functional code to an imperative language like OpenCL or CUDA in a way that preserves our chosen strategy.
This paper presents a formal approach to solving this translation problem.

\subsection{Strategy Preserving Translation to Imperative Code}

What is the simplest way of converting a functional program to an imperative one?  Starting with our zip-map-reduce formulation of dot-product (\ref{code-ex:simple-dot-product}), we can turn it into an imperative program simply by assigning its result to an output variable $\mathit{out}$:
\begin{displaymath}
  \mathit{out} := \prim{reduce}~(+)~0~(\prim{map}~(\lambda x.~\prim{fst}~x*\prim{snd}~x)~(\prim{zip}~\mathit{xs}~\mathit{ys}))
\end{displaymath}
Unfortunately, this is not suitable for compilation targets like OpenCL or CUDA.
While assignment statements are the bread-and-butter of such languages, their expression languages certainly do not include such modern amenities as higher order $\prim{map}$ and $\prim{reduce}$ functions.
To translate these away, we introduce a novel \emph{acceptor-passing} translation $\semA{E}_\delta(\mathit{out})$.
The key idea is that for any expression $E$ producing data of type $\delta$, the translation $\semA{E}_\delta(\mathit{out})$ is an imperative program that has the same effect as the assignment $\mathit{out} := E$ and is free from higher-order combinators.
This translation is mutually defined with a continuation passing translation $\semE{E}_\delta(C)$ that takes a parameterised command $C$ that will consume the output, instead of taking an output variable.

The definition of the translation is given in \autoref{sec:translation-i}.
We introduce it here by example.
Applied to our dot-product code, our translation first replaces the $\prim{reduce}$ by a corresponding imperative combinator $\prim{reduceI}$.
We will see below that $\prim{reduceI}$ is straightforwardly implemented in terms of variable allocation and a for-loop.
\begin{displaymath}
  \begin{array}{cl}
    & \semA{\prim{reduce}~(+)~0~(\prim{map}~(\lambda x.~\prim{fst}~x*\prim{snd}~x)~(\prim{zip}~\mathit{xs}~\mathit{ys}))}_{\mathsf{num}}(\mathit{out}) \\[.25em]
    =&\begin{array}[t]{@{}l}
        \semE{\prim{map}~(\lambda x.~\prim{fst}~x*\prim{snd}~x)~(\prim{zip}~\mathit{xs}~\mathit{ys})}_{n.\mathsf{num}}(\lambda x.~\\
        \quad \semE{0}_{\mathsf{num}}(\lambda y.~
        \prim{reduceI}~n~(\lambda x~y~o.~\semA{x + y}_{\mathsf{num}}(o))~y~x~(\lambda r.~\semA{r}(A))))
      \end{array}
  \end{array}
\end{displaymath}
The $\prim{map}$ is now translated, by the continuation passing translation, into allocation of a temporary array and an imperative $\prim{mapI}$ combinator.
As with $\prim{reduceI}$, the $\prim{mapI}$ combinator is straightforwardly implementable in terms of a (parallel) for-loop.
The operator $\prim{new}~\delta~\mathit{ident}$ declares a new storage cell of type $\delta$ named $\mathit{ident}$, where storage cells are represented as pairs of an acceptor (i.e., ``writer'', ``l-value'') part $\mathit{ident}.1$ and an expression (i.e. ``reader'', ``r-value'') part $\mathit{ident}.2$.
Our language, which we introduce in \autoref{sec:typeSystem}, is a variant of Reynolds' Idealised Algol \citep{Reynolds78}.
\begin{displaymath}
  \begin{array}{cl}
    =&\prim{new}~(n.\mathsf{num})~(\lambda \mathit{tmp}.~
      \begin{array}[t]{@{}l}
        \semE{\prim{zip}~\mathit{xs}~\mathit{ys}}_{n.(\mathsf{num}\times\mathsf{num})}(\lambda x.~
           \prim{mapI}~n~(\lambda x~o.~\semA{\prim{fst}~x * \prim{snd}~x}_{\mathsf{num}}(o))~x~\mathit{tmp}.1); \\
         \semE{0}_{\mathsf{num}}(\lambda y.~
           \prim{reduceI}~n~(\lambda x~y~o.~\semA{x + y}_{\mathsf{num}}(o))~y~\mathit{tmp}.2~(\lambda r.~\semA{r}(A))))
      \end{array}
  \end{array}
\end{displaymath}
Readers familiar with other translations of data parallel functional programs into imperative loops may be surprised at the allocation of a temporary array here.
Typically, the compilation process would be expected to automatically fuse the computation of the $\prim{map}$ into the translation of the $\prim{reduce}$.
However, this is precisely what we do \emph{not} want from a predictable compilation process for parallelism.
If fusion is desired, it is carried out before this translation is applied and directly encoded in the functional program, as seen earlier in example~(\ref{code-ex:dot-product-complex}).
The parallelism strategy described by the functional code here precisely states ``parallel map, followed by sequential reduce''.
Predictability of the translation is essential for more complex parallelism strategies that exploit parallelism hierarchies and even different memory address spaces as we will see later in \autoref{sec:opencl:addrspace}.

Continuing the translation process, we substitute $\mathit{out}$, the arithmetic expressions and the $\prim{zip}$, leaving two uses of the ``intermediate-level'' combinators $\prim{mapI}$ and $\prim{reduceI}$:
\begin{displaymath}
  \begin{array}{cl}
    =&\prim{new}~(n.\mathsf{num})~(\lambda \mathit{tmp}.~
        \begin{array}[t]{@{}l}
          \prim{mapI}~n~(\lambda x~o.~o := \prim{fst}~x * \prim{snd}~x)~(\prim{zip}~\mathit{xs}~\mathit{ys})~\mathit{tmp}.1; \\
          \prim{reduceI}~n~(\lambda x~y~o.~o := x + y)~0~\mathit{tmp}.2~(\lambda r.~\mathit{out} := r)) \\
        \end{array}
  \end{array}
\end{displaymath}
These combinators are now replaced by parallel and sequential for-loops, which we describe in a further translation stage in \autoref{sec:translation-ii}.
\begin{smallequation}\label{codeex:dpia-dot-product}
  \begin{array}{cl}
    =&\prim{new}~(n.\mathsf{num})~(\lambda \mathit{tmp}.~\begin{array}[t]{@{}l}
        \prim{parfor}~n~\mathit{tmp}.1~(\lambda i~o.~
          o := \prim{fst}~(\prim{idx}~(\prim{zip}~\mathit{xs}~\mathit{ys})~i) * \prim{snd}~(\prim{idx}~(\prim{zip}~\mathit{xs}~\mathit{ys})~i)); \\
        \prim{new}~\mathsf{num}~(\lambda \mathit{accum}.~
          \begin{array}[t]{@{}l}
            \mathit{accum}.1 := 0;\\
            \prim{for}~n~(\lambda i.~\mathit{accum}.1 := \mathit{accum}.2 + \prim{idx}~\mathit{tmp}.2~i); \\
            \mathit{out} := \mathit{accum.2})) \\
         \end{array} \\
      \end{array} \\
  \end{array}
\end{smallequation}
The sequential $\prim{for}$ loops of our intermediate language are standard; $\prim{for}~n~(\lambda i.~b)$ executes the body $b$ $n$ times with iteration counter $i$. Parallel $\prim{parfor}$ loops are slightly more complex due to the way they explicitly take a parameter (here named $\mathit{tmp}.1$) that describes where to place the results of each iteration in a data-race free way. We describe this fully in \autoref{sec:dpia-prims}.

We are now left with an imperative program, albeit with a non-standard parallel-for construct and complex data access expressions involving $\prim{fst}$, $\prim{zip}$, $\prim{idx}$ and so on. Our final translation to pseudo-C (\autoref{sec:translation-iii}) resolves these data layout expressions into explicit indexing computations:
\begin{lstlisting}
float tmp[N];
parfor (int i = 0; i < N; i += 1)
  tmp[i] = xs[i] * ys[i];
float accum = 0.0;
for (int i = 0; i < N; i += 1)
  accum = accum + tmp[i];
output = accum;
\end{lstlisting}
This resulting low level imperative code precisely implements the strategy ``parallel map, followed by sequential reduce'' described by our original functional expression (\ref{code-ex:simple-dot-product}).

Our original dot-product code does not produce particularly complex code, but our translation method scales to more detailed parallelism strategies. The alternative dot-product code in (\ref{code-ex:dot-product-complex}), which rearranges the $\prim{map}$ and $\prim{reduce}$ combinators in order to better exploit parallel hardware, yields the following code:
\begin{lstlisting}
float tmp[N/2048];
parfor (int i = 0; i < N/(2048*128); i += 1) {
  parfor (int j = 0; j < 128; j += 1) {
    float accum = 0.0;
    for (int k = 0; k < 2048; k += 1) {
      accum = (xs[(2048*128 * i) + (128 * j) + k]
             * ys[(2048*128 * i) + (128 * j) + k]) + accum; }
    tmp[((128 * i) + j)] = accum;
  }
}
float accum = 0.0;
for (int i = 0; i < N/2048; i += 1) {
  accum = accum + tmp[i];
}
output = accum;
\end{lstlisting}
As we shall see in \autoref{sec:experimentalResults}, given a target-architecture optimised parallelisation strategy defined in functional code, our translation process produces OpenCL code with performance on a par with previous ad hoc code generators, and with hand written code.

Key to our translation methodology is a single intermediate language that can express pure functional expressions and deterministic race free parallel imperative programs, and which is amenable to formal reasoning. In the next section, we describe our language for this task, $\DPIA$: \emph{Data Parallel Idealised Algol}.

\section{Data Parallel Idealised Algol}
\label{sec:typeSystem}

Our intermediate language for code generation is an extension of Reynolds' \emph{Idealised Algol} \citep{Reynolds97}. Idealised Algol is the orthogonal combination of typed $\lambda$-calculus and imperative programming. To support deterministic race free parallel programming, we use the ``Syntactic Control of Interference Revisited'' variant of Idealised Algol \citep{OHearnPTT99}, extended with size-indexed array and tuple data types. We call our language \emph{Data Parallel Idealised Algol} (\DPIA). We saw examples of \DPIA in the previous section. We now highlight the major features of \DPIA, and then give the formal presentation of its types (\autoref{sec:types}), type system (\autoref{sec:typing-rules}), and data parallel programming primitives (\autoref{sec:dpia-prims}). We discuss the formal semantics of \DPIA in \autoref{sec:correctness}.

\paragraph{Orthogonal Combination of Typed $\lambda$-Calculus and Imperative Programming}
The central feature of the Idealised Algol family of languages is the orthogonal combination of an imperative language with a typed $\lambda$-calculus. Unlike traditional functional languages like Scheme, Haskell and ML, $\beta$-reduction is not the main engine of computation; for example, there is no pattern matching. The purpose of $\lambda$-abstraction is to add a facility for procedural abstraction to a base imperative language. In the absence of recursion in the $\lambda$-calculus component of the language, given a whole program it is possible to normalise away all $\lambda$-abstractions to yield a ``pure'' imperative program. The $\lambda$-calculus therefore becomes almost a meta-language for constructing imperative programs. We will exploit this feature during our translation process in \autoref{sec:translation}.

\paragraph{Substructural Types for Interference Control}
\citet{Reynolds78} notes that the introduction of a procedure facility into an imperative programming language destroys a nice property of imperative programs without procedures: that distinct identifiers always refer to distinct parts of the store. Such aliasing complicates reasoning, especially the reasoning required to show that running code in parallel is deterministic and data race free. To re-enable straightforward reasoning in the presence of procedures, Reynolds introduced a discipline, \emph{Syntactic Control of Interference} (SCI), that ensures that distinct identifiers never interfere. This is more subtle than it may first appear, due to identifiers that are used \emph{passively} (essentially read-only), which are allowed to alias. We build upon the \emph{Syntactic Control of Interference Revisited} (SCIR) system introduced by \citet{OHearnPTT99}, which presents Reynolds ideas as a substructural typed $\lambda$-calculus (see \autoref{fig:typing-rules}). Type based interference control ensures, by construction, that we have sufficient information to guarantee data race freedom in our generated code.

\paragraph{Primitives for Data Parallel Programming}
Idealised Algol, with or without Syntactic Control of Interference, has primarily been studied theoretically as a core calculus combining imperative programming with $\lambda$-abstraction (with the notable exception of Ghica's use of SCI for hardware synthesis \citep{Ghica07}). To use SCIR as an intermediate language for data parallel computation, we extend it with compound data types, tuples and arrays, and with indexed types to account for array size polymorphism and data type polymorphism. We also extend SCIR with primitives designed for data parallel programming. The central primitive is a data race free parallel for-loop primitive, which we describe in \autoref{sec:dpia-prims}.

\subsection{The Types of \DPIA}
\label{sec:types}

The type system of \DPIA, following Idealised Algol, separates \emph{data types}, which classify data (integers, floats, arrays, \emph{etc.}), from \emph{phrase types}, which classify the parts of a program according to the interface they offer. Phrase types are a generalisation to first-class status of the syntactic categories in a standard imperative language that distinguish between expressions (r-values), l-values, and statements. Phrase types in \DPIA comprise \emph{expressions}, which produce data, possibly reading from the store; \emph{acceptors}, which describe modifiable areas of the store (analogous to \emph{l-values} in imperative languages~\citep{Strachey00}); \emph{commands}, which modify the store; \emph{functions}, which are parameterised phrases; and \emph{records}, which offer a choice of multiple phrases. The separation into data and phrase types distinguishes Idealised Algol-style type systems from those for functional languages, which commonly use expression phrases for everything (permitting, for example, functional data).

To facilitate interference control, we identify a subset of phrase types which are \emph{passive} (\autoref{sec:passive-types}), i.e.~essentially read-only, and so are safe to share across parallel threads. (We elaborate on what ``essentially read-only'' means in \autoref{sec:passive-types} and \autoref{sec:typing-rules}.)

\subsubsection{Kinding rules}
\label{sec:kinding-rules}

\begin{figure}[t]
  \small
  \begin{minipage}{0.3\linewidth}
    \begin{mathpar}
      \kappa ::= \datatype \mid \phrasetype \mid \natkind
    \end{mathpar}
    \subcaption{Kinds}\label{fig:kinds}
  \end{minipage}%
  \begin{minipage}{0.3\linewidth}
    \begin{mathpar}
      \inferrule*
      {x : \kappa \in \Delta}
      {\Delta \vdash x : \kappa}
    \end{mathpar}
    \subcaption{Kinding Structural Rules}\label{fig:structural-kinding}
  \end{minipage}
  \begin{minipage}{0.3\linewidth}
    \begin{mathpar}
       \inferrule
       {\models \forall \sigma : \mathit{dom}(\Delta) \to \mathbb{N}.\sigma(I) = \sigma(J)}
       {\Delta \vdash I \equiv J : \natkind}
    \end{mathpar}
    \subcaption{Type Equality}\label{fig:equality-kinding}
  \end{minipage}

  \medskip

  \begin{minipage}{1.0\linewidth}
    \begin{mathpar}
      \inferrule*
      { }
      {\Delta \vdash \underline{n} : \natkind}

      \inferrule*
      {\Delta \vdash I : \natkind \\ \Delta \vdash J : \natkind}
      {\Delta \vdash I + J : \natkind}

      \inferrule*
      {\Delta \vdash I : \natkind \\ \Delta \vdash J : \natkind}
      {\Delta \vdash I J : \natkind}
    \end{mathpar}
    \subcaption{Natural numbers}\label{fig:natural-number-kinding}
  \end{minipage}

  \medskip

  \begin{minipage}{1.0\linewidth}
    \begin{mathpar}
      \inferrule*
      { }
      {\Delta \vdash \mathsf{num} : \datatype}

      \inferrule*
      {\Delta \vdash I : \natkind}
      {\Delta \vdash \prim{idx}(I) : \datatype}

      \inferrule*
      {\Delta \vdash I : \natkind \\\\ \Delta \vdash \delta : \datatype}
      {\Delta \vdash I.\delta : \datatype}

      \inferrule*
      {\Delta \vdash \delta_1 : \datatype \\\\ \Delta \vdash \delta_2 : \datatype}
      {\Delta \vdash \delta_1 \times \delta_2 : \datatype}
    \end{mathpar}
    \subcaption{Data Types}\label{fig:data-type-kinding}
  \end{minipage}

  \medskip

  \begin{minipage}{1.0\linewidth}
    \begin{mathpar}
      \inferrule*
      {\Delta \vdash \delta : \datatype}
      {\Delta \vdash \tyexp[\delta] : \phrasetype}

      \inferrule*
      {\Delta \vdash \delta : \datatype}
      {\Delta \vdash \tyacc[\delta] : \phrasetype}

      \inferrule*
      { }
      {\Delta \vdash \tycomm : \phrasetype}

      \inferrule*
      {\Delta \vdash \theta_1 : \phrasetype \\
        \Delta \vdash \theta_2 : \phrasetype}
      {\Delta \vdash \theta_1 \times \theta_2 : \phrasetype}

      \inferrule*
      {\Delta \vdash \theta_1 : \phrasetype \\\\
        \Delta \vdash \theta_2 : \phrasetype}
      {\Delta \vdash \theta_1 \to \theta_2 : \phrasetype}

      \inferrule*
      {\Delta \vdash \theta_1 : \phrasetype \\\\
        \Delta \vdash \theta_2 : \phrasetype}
      {\Delta \vdash \theta_1 \pureto \theta_2 : \phrasetype}

      \inferrule*
      {\Delta, x : \kappa \vdash \theta : \phrasetype \\ \kappa \in \{\natkind,\datatype\}}
      {\Delta \vdash (x\mathord:\kappa) \to \theta : \phrasetype}
    \end{mathpar}
    \subcaption{Phrase Types}\label{fig:phrase-type-kinding}
  \end{minipage}

  \caption{Well-formed Types}
  \label{fig:types}
\end{figure}

We extend SCIR with both data type and size polymorphism, so we need a kind system. \autoref{fig:types} presents the kinding rules for \DPIA types. The kinds $\kappa$ of \DPIA include the major classifications into data types ($\datatype$) and phrase types ($\phrasetype$), along with the kind of type-level natural numbers ($\natkind$). Types may contain variables, so we use a kinding judgement $\Delta \vdash \tau : \kappa$, which states that type $\tau$ has kind $\kappa$ in kinding context $\Delta$. \autoref{fig:structural-kinding} gives the variable rule that permits the use of type variables in well-kinded types. \autoref{fig:natural-number-kinding} presents the rules for type-level natural numbers: either constants $\underline{n}$, addition $I + J$, or multiplication $I J$ (where $I$ and $J$ range over terms of kind $\natkind$).

The rules for data types are presented in \autoref{fig:data-type-kinding}. We use $\delta$ to range over data types. The base types are $\mathsf{num}$ for numbers; and a data type of array indexes $\prim{idx}(n)$, parameterised by the maximum array index. There are two compound types of data. For any data type $\delta$ and natural number term $I$, $I.\delta$ is the data type of homogeneous arrays of $\delta$s of size $I$. (We opt for a concise notation for array types as they are pervasive in data parallel programming.) Heterogeneous compound data types (records) are built using the rule for $\delta_1 \times \delta_2$.

The phrase types of \DPIA are given in \autoref{fig:phrase-type-kinding}. We use $\theta$ to range over phrase types. For each data type $\delta$, there are phrase types $\mathsf{exp}[\delta]$ for \emph{expression} phrases that produce data of type $\delta$, and $\mathsf{acc}[\delta]$ for \emph{acceptor} phrases that consume data of type $\delta$. The $\mathsf{comm}$ phrase type classifies \emph{command} phrases that may modify the store. Phrases that can be used in two different ways, $\theta_1$ or $\theta_2$, are classified using the phrase product type $\theta_1 \times \theta_2$. This type is distinct from the \emph{data} product type $\delta_1 \times \delta_2$: the data type represents a pair of data values; the phrase type represents an ``interface'' that offers two possible ``methods''. (For readers familiar with Linear Logic \cite{DBLP:journals/tcs/Girard87}, the phrase product is like ``with'' ($\with$) and the data product like ``tensor'' $(\otimes)$.) The final three phrase types are all variants of parameterised phrase types. The phrase types $\theta_1 \to \theta_2$ and $\theta_1 \pureto \theta_2$ classify phrase functions. The $\mathrm{p}$ subscript denotes passive functions. The phrase type $(x\mathord:\kappa) \to \theta$ classifies a phrase that is parameterised either by a data type or a natural number.

The types of \DPIA include arithmetic expressions, so we have a non trivial notion of equality between types, written $\Delta \vdash \tau_1 \equiv \tau_2 : \kappa$. The key type equality rule is given in \autoref{fig:equality-kinding}: two arithmetic expressions are equal if they are equal as natural numbers for all interpretations ($\sigma$) of their free variables. This equality is lifted to all other types by structural congruence.

\subsubsection{Passive Types}
\label{sec:passive-types}

\autoref{fig:passive-types} identifies the subset of phrase types that classify passive phrases. The opposite of passive is \emph{active}. We use $\phi$ to range over passive phrase types. An expression phrase type $\mathsf{exp}[\delta]$ is always passive --- phrases of this type can, by definition, only read the store. A compound phrase type is always passive if its component phrase types are all passive. Furthermore, a passive function type $\theta_1 \pureto \theta_2$ is always passive, and a plain function type is passive whenever its return type is passive (irrespective of the argument type).

Passive types are essentially read-only. The one exception whereby a phrase of passive type may modify the store is a passive function with active argument and return types. Such a function can only modify the part of the store addressable through the active phrase it is supplied with as an argument.

\begin{figure}[t]
  \small
  \centering
  \begin{mathpar}
    \inferrule*
    {\Delta \vdash \delta : \datatype}
    {\Delta \vdash \tyexp[\delta] : \passivetype}

    \inferrule*
    {\Delta \vdash \phi_1 : \passivetype \\
      \Delta \vdash \phi_2 : \passivetype}
    {\Delta \vdash \phi_1 \times \phi_2 : \passivetype}

    \inferrule*
    {\Delta \vdash \theta : \phrasetype \\
      \Delta \vdash \phi : \passivetype}
    {\Delta \vdash \theta \to \phi : \passivetype}

    \inferrule*
    {\Delta \vdash \theta_1 : \phrasetype \\
      \Delta \vdash \theta_2 : \phrasetype}
    {\Delta \vdash \theta_1 \pureto \theta_2 : \passivetype}

    \inferrule*
    {\Delta, x : \kappa \vdash \theta : \passivetype \\ \kappa \in \{\natkind,\datatype\}}
    {\Delta \vdash (x\mathord:\kappa) \to \theta : \passivetype}
  \end{mathpar}
  \caption{Passive Types}
  \label{fig:passive-types}
\end{figure}

\subsection{Typing Rules for \DPIA}
\label{sec:typing-rules}

The typing judgement of \DPIA follows the SCIR system of \citet{OHearnPTT99} in distinguishing between passive and active uses of identifiers. Our judgement also has a kinding context for size and data type polymorphism. The judgement form has the following structure:
\begin{displaymath}
  \typd{\Pi}{\Gamma}{P}{\theta}
\end{displaymath}
where $\Delta$ is the kinding context, $\Pi$ is a context of passively used identifiers, $\Gamma$ is a context of actively used identifiers, $P$ is a program phrase, and $\theta$ is a phrase type. All the types in $\Pi$ and $\Gamma$ are phrase types well-kinded by $\Delta$. The phrase type $\theta$ must also be well-kinded by $\Delta$. The order of entries does not matter in any of the contexts. The contexts $\Delta$ and $\Pi$ are subject to contraction and weakening; context $\Gamma$ is not.

The split context formulation of SCIR recalls that of Barber's DILL system \citep{barber96dual}, which also distinguishes between linear and unrestricted assumptions. The SCIR system differs in how movement between the zones is mediated in terms of passive and active types. Section 2.6 of \citep{OHearnPTT99} discusses the relationship between SCIR and Linear Logic.

The core typing rules of \DPIA are given in \autoref{fig:typing-rules}. These rules define how variable phrases are formed, how parameterised and compound phrases are introduced and eliminated, and how passive and active types are managed. Any particular application of \DPIA is specified by giving a collection of primitive phrases \textsc{Primitives}, each of which has a closed phrase type. We describe a collection for data parallel programming in \autoref{sec:dpia-prims}.

\autoref{fig:structural-rules} presents the rule for forming variable phrases, implicit conversion between equal types, and the use of primitives. At point of use, all variables are considered to be used actively. If the final phrase type is passive, then an active use may be converted to a passive one by the \TirName{Passify} rule. Primitives may be used in any context. \autoref{fig:intro-elim-rules} presents the rules for parameterised and compound phrases. These are all standard typed $\lambda$-calculus style rules, except the use of separate contexts for a function and its arguments in the \TirName{App} rule. This ensures that every function and its argument use non-interfering active resources, maintaining the invariant that distinct identifiers refer to non-interfering phrases. Note that we do not require separate contexts for the two parts of a compound phrase in the \TirName{Pair} rule. Compound phrases offer two ways of interacting with the \emph{same} underlying resource (as in the with ($\with$) rule from linear logic).

\autoref{fig:active-passive-rules} describes how passive and active uses of variables are managed. The \TirName{Activate} rule allows any variable that has been used passively to be treated as if it were used actively. The \TirName{Passify} rule allows active uses to be treated as passive, as long as the final phrase type is passive. The \TirName{Promote} rule turns functions into passive functions, as long as they do not contain any free variables used actively. The \TirName{Derelict} rule indicates that a passive function can always be seen as a normal function, if required.

\begin{figure*}[t]
  \small
  \begin{minipage}{1.0\linewidth}
    \begin{mathpar}
      \inferrule* [right=Var]
      { }
      {\typd{\Pi}{\Gamma, x : \theta, \Gamma'}{x}{\theta}}

      \inferrule* [right=Conv]
      {\typd{\Pi}{\Gamma}{P}{\theta_1} \\\\
        \Delta \vdash \theta_1 \equiv \theta_2 : \phrasetype}
      {\typd{\Pi}{\Gamma}{P}{\theta_2}}

      \inferrule* [right=Prim]
      {\mathsf{prim} : \theta \in \textsc{Primitives}}
      {\typd{\Pi}{\Gamma}{\mathsf{prim}}{\theta}}
    \end{mathpar}
    \subcaption{Structural Rules}\label{fig:structural-rules}
  \end{minipage}

  \medskip

  \begin{minipage}{1.0\linewidth}
    \begin{mathpar}
      \inferrule* [right=Lam]
      {\typd{\Pi}{\Gamma, x : \theta_1}{P}{\theta_2}}
      {\typd{\Pi}{\Gamma}{\lambda x.P}{\theta_1 \to \theta_2}}

      \inferrule* [right=App]
      {\typd{\Pi}{\Gamma_1}{P}{\theta_1 \to \theta_2} \\
        \typd{\Pi}{\Gamma_2}{Q}{\theta_1}}
      {\typd{\Pi}{\Gamma_1, \Gamma_2}{P~Q}{\theta_2}}\medskip
    \end{mathpar}
    \begin{mathpar}
      \inferrule* [right=TLam]
      {\typ{\Delta, x : \kappa}{\Pi}{\Gamma}{P}{\theta} \\
        x \not\in \mathit{fv}(\Pi, \Gamma)}
      {\typd{\Pi}{\Gamma}{\Lambda x. P}{(x \mathord: \kappa) \to \theta}}

      \inferrule* [right=TApp]
      {\typd{\Pi}{\Gamma}{P}{(x \mathord:\kappa) \to \theta} \\
        \Delta \vdash e : \kappa}
      {\typd{\Pi}{\Gamma}{P~e}{\theta[e/x]}}\medskip
    \end{mathpar}
    \begin{mathpar}
      \inferrule* [right=Pair]
      {\typd{\Pi}{\Gamma}{P}{\theta_1} \\
        \typd{\Pi}{\Gamma}{Q}{\theta_2}}
      {\typd{\Pi}{\Gamma}{\langle P, Q \rangle}{\theta_1 \times \theta_2}}

      \inferrule* [right=Proj]
      {\typd{\Pi}{\Gamma}{P}{\theta_1 \times \theta_2}}
      {\typd{\Pi}{\Gamma}{P.i}{\theta_i}}
    \end{mathpar}
    \subcaption{Introduction and Elimination Rules}\label{fig:intro-elim-rules}
  \end{minipage}

  \medskip

  \begin{minipage}{1.0\linewidth}
    \begin{mathpar}
      \inferrule* [right=Activate]
      {\typd{\Pi, x : \theta}{\Gamma}{P}{\theta'}}
      {\typd{\Pi}{\Gamma, x : \theta}{P}{\theta'}}

      \inferrule* [right=Passify]
      {\typd{\Pi}{\Gamma, x : \theta}{P}{\phi}}
      {\typd{\Pi, x : \theta}{\Gamma}{P}{\phi}}\medskip
    \end{mathpar}
    \begin{mathpar}
      \inferrule* [right=Promote]
      {\typd{\Pi}{\cdot}{P}{\theta_1 \to \theta_2}}
      {\typd{\Pi}{\cdot}{P}{\theta_1 \pureto \theta_2}}

      \inferrule* [right=Derelict]
      {\typd{\Pi}{\Gamma}{P}{\theta_1 \pureto \theta_2}}
      {\typd{\Pi}{\Gamma}{P}{\theta_1 \to \theta_2}}
    \end{mathpar}
    \subcaption{Active and Passive Phrase Rules}\label{fig:active-passive-rules}
  \end{minipage}

  \caption{Typing Rules: Indexed Affine Linear $\lambda$-Calculus with Passivity \cite{OHearnPTT99}}
  \label{fig:typing-rules}
\end{figure*}

\paragraph{\DPIA's functional sub-language} By inspection of the rules, we can see that if we restrict to phrase types constructed from $\tyexp[\delta]$, functions, polymorphic functions, and tuples, then the constraints on multiple uses of variables in \DPIA cease to apply. Therefore, \DPIA has a sub-language that has the same type system as a normal (non-substructural) typed $\lambda$-calculus with base types for numbers, arrays and tuples, and a limited form of polymorphism. When we introduce the functional primitives for \DPIA in the next section, we will enrich this $\lambda$-calculus with arithmetic, array manipulators, and higher-order array combinators. It is this purely functional sub-language of \DPIA that allows us to embed functional data parallel programs in a semantics preserving way.

\subsection{Data Parallel Programming Primitives}
\label{sec:dpia-prims}

\begin{figure}
  \small
  \begin{minipage}{1.0\linewidth} \begin{tabular*}{\linewidth}{>{$}l<{$}@{\hspace{0.4em}}>{$}c<{$}>{$}l<{$}}
        \underline{n}&:&\tyexp[\mathsf{num}] \\
        \prim{negate}&:&\tyexp[\mathsf{num}] \to \tyexp[\mathsf{num}] \\
        (+,*,/,-)   &:&\tyexp[\mathsf{num}] \times \tyexp[\mathsf{num}] \to \tyexp[\mathsf{num}] \bigskip\\

        \prim{map}&:&(n : \natkind) \to (\delta_1~\delta_2 : \datatype) \to (\tyexp[\delta_1] \to \tyexp[\delta_2]) \to \tyexp[n.\delta_1] \to \tyexp[n.\delta_2] \\
        \prim{reduce}&:&(n : \natkind) \to (\delta_1~\delta_2 : \datatype) \to 
        (\tyexp[\delta_1] \to \tyexp[\delta_2] \to \tyexp[\delta_2]) \to \tyexp[\delta_2] \to \tyexp[n.\delta_1] \to \tyexp[\delta_2]\bigskip\\

        \prim{zip}&:&(n : \natkind) \to (\delta_1~\delta_2 : \datatype) \to \tyexp[n.\delta_1] \to \tyexp[n.\delta_2] \to \tyexp[n.(\delta_1 \times \delta_2)] \\
        \prim{split}&:&(n~m : \natkind) \to (\delta : \datatype) \to \tyexp[nm.\delta] \to \tyexp[m.n.\delta] \\
        \prim{join}&:&(n~m : \natkind) \to (\delta : \datatype) \to \tyexp[n.m.\delta] \to \tyexp[nm.\delta] \\
        \prim{pair}&:&(\delta_1~\delta_2: \datatype) \to \tyexp[\delta_1] \to \tyexp[\delta_2] \to \tyexp[\delta_1 \times \delta_2]\\
        \prim{fst}&:&(\delta_1~\delta_2 : \datatype) \to \tyexp[\delta_1 \times \delta_2] \to \tyexp[\delta_1] \\
        \prim{snd}&:&(\delta_1~\delta_2 : \datatype) \to \tyexp[\delta_1 \times \delta_2] \to \tyexp[\delta_2] \\
    \end{tabular*}
    \subcaption{Functional primitives}\label{fig:func-prim}
  \end{minipage}

  \vspace{1em}

  \begin{minipage}{1.0\linewidth}
  \begin{tabular*}{\linewidth}{>{$}l<{$}>{$}c<{$}>{$}l<{$}}
        \prim{skip}&:&\tycomm \\
        (\mathord;)&:&\tycomm \times \tycomm \to \tycomm \\
        \prim{new}&:&(\delta : \datatype) \to (\tyvar[\delta] \to \tycomm) \to \tycomm
        \qquad (\text{where }\mathsf{var}[\delta] = \mathsf{acc}[\delta] \times \mathsf{exp}[\delta] : \phrasetype) \\
        (:=)&:&\tyacc[\mathsf{num}] \times \tyexp[\mathsf{num}] \to \tycomm \\
        \prim{for}&:&(n : \natkind) \to (\tyexp[\mathrm{idx}(n)] \to \tycomm) \to \tycomm \\
        \prim{parfor}&:&(n : \natkind) \to (\delta : \datatype) \to \tyacc[n.\delta] \to (\tyexp[\mathrm{idx}(n)] \to \tyacc[\delta] \pureto \tycomm) \to \tycomm \bigskip\\

        \prim{splitAcc}&:&(n~m : \natkind) \to (\delta : \datatype) \to \tyacc[m.n.\delta] \to \tyacc[nm.\delta] \\
        \prim{joinAcc}&:&(n~m : \natkind) \to (\delta : \datatype) \to \tyacc[nm.\delta] \to \tyacc[n.m.\delta] \\
        \prim{pairAcc_1}&:&(\delta_1~\delta_2 : \datatype) \to \tyacc[\delta_1 \times \delta_2] \to \tyacc[\delta_1] \\
        \prim{pairAcc_2}&:&(\delta_1~\delta_2 : \datatype) \to \tyacc[\delta_1 \times \delta_2] \to \tyacc[\delta_2] \\
        \prim{zipAcc_1} &:& (n : \natkind) \to (\delta_1 \delta_2 : \datatype) \to \tyacc[n.\delta_1 \times \delta_2] \to \tyacc[n.\delta_1] \\
        \prim{zipAcc_2} &:& (n : \natkind) \to (\delta_1 \delta_2 : \datatype) \to \tyacc[n.\delta_1 \times \delta_2] \to \tyacc[n.\delta_2] \bigskip\\

        \prim{idx} &:&(n : \natkind) \to (\delta : \datatype) \to \tyexp[n.\delta] \to \tyexp[\mathrm{idx}(n)] \to \tyexp[\delta] \\
        \prim{idxAcc} &:&(n : \natkind) \to (\delta : \datatype) \to \tyacc[n.\delta] \to \tyexp[\mathrm{idx}(n)] \to \tyacc[\delta]
    \end{tabular*}
    \subcaption{Imperative primitives}\label{fig:imp-prim}
  \end{minipage}

  \vspace{2em}

  \begin{minipage}{1.0\linewidth}
    \begin{tabular*}{\linewidth}{>{$}l<{$}@{\hspace{1.4em}}>{$}c<{$}>{$}l<{$}}
        \prim{mapI}   &:& (n : \natkind) \to (\delta_1~\delta_2 : \datatype) \to
                            (\tyexp[\delta_1] \to \tyacc[\delta_2] \pureto \tycomm) \to \tyexp[n.\delta_1] \to \tyacc[n.\delta_2] \to \tycomm \\
        \prim{reduceI}&:& (n: \natkind) \to (\delta_1~\delta_2 : \datatype) \to (\tyexp[\delta_1] \to \tyexp[\delta_2] \to \tyacc[\delta_2] \to \tycomm) \to \\
                        & & \qquad \tyexp[\delta_2] \to \tyexp[n.\delta_1] \to (\tyexp[\delta_2] \to \tycomm) \to \tycomm \\
    \end{tabular*}
    \subcaption{Intermediate imperative combinators}\label{fig:imp-intermediate}
  \end{minipage}

  \vspace{1em}

  \caption{Data Parallel Programming Primitives, Functional and Imperative}
  \label{fig:primitives}
\end{figure}

\autoref{sec:typing-rules} has described a general framework for a language with interference control. We now instantiate this framework with typed primitive operations for data parallel programming, outlined in \autoref{fig:primitives}. Our primitives fall into two principal categories: high-level functional primitives, and low-level imperative primitives. Programs that are the input to our translation process are composed of the high-level functional primitives. These programs contain uses of $\prim{map}$ and $\prim{reduce}$ that have no counterpart in low-level languages for data-parallel computation. Our translation process converts these into low-level combinators (\autoref{sec:translation-i}). A final lowering translation removes all functional primitives except arithmetic (\autoref{sec:translation-iii}).

As primitives are treated specially by the \textsc{Prim} rule they can (with the aid of a little $\eta$-expansion) always be promoted to be passive. Thus, it is never necessary to annotate the arrows of a first-order primitive with a $\mathrm{p}$ subscript. The only such annotations that are necessary are those final arrows of function types occurring inside the type of a higher-order primitive that is required to be passive (in our case, only $\prim{parfor}$ and $\prim{mapI}$).

\paragraph{Functional Primitives}

\autoref{fig:func-prim} lists the type signatures of the primitives used for constructing purely functional data parallel programs. These fit into three groups. The first group consists of numeric literals ($\underline{n}$) and first-order operations on scalars ($\prim{negate},(+),(-),(*),(/)$).
The second group contains the two key higher-order functional combinators for constructing array processing programs: $\prim{map}$ and $\prim{reduce}$. These have (the Idealised Algol renditions of) the standard types for these primitives, extended with size information. The third group comprises functions for manipulating data layouts: $\prim{zip}$ joins two arrays of equal length into an array of pairs, $\prim{split}$ breaks a one dimensional array into a two dimensional array and $\prim{join}$ flattens a two dimensional array into a one dimensional array, $\prim{pair}$ constructs a pair, and $\prim{fst}$ and $\prim{snd}$ deconstruct a pair. All of these primitives are data type indexed and those that operate on arrays are also size indexed.

\paragraph{Example: dot-product}
The dot-product example from \autoref{sec:motivation} is written using the functional primitives like so, for input vectors $\mathit{xs}$ and $\mathit{ys}$ of length $n$, the only difference being that all of the size and data type information is described in detail (often we can infer these arguments and so in practice we often omit them):
\begin{displaymath}
  \prim{reduce}~n~\mathsf{num}~\mathsf{num}~
  (\lambda x~y.~x+y)~
  \underline{0}~
  (\prim{map}
  \begin{array}[t]{@{\hspace{0.3em}}l}
    n~(\mathsf{num} \times \mathsf{num})~\mathsf{num} \\
    (\lambda x.~\prim{fst}~\mathsf{num}~\mathsf{num}~x * \prim{snd}~\mathsf{num}~\mathsf{num}~x)\\
    (\prim{zip}~n~\mathsf{num}~\mathsf{num}~\mathit{xs}~\mathit{ys}))
  \end{array}
\end{displaymath}
Likewise, the specialised version of dot-product from \autoref{sec:motivation} with nested $\prim{split}$s and $\prim{join}$s can be expressed with detailed size and type information throughout.

\paragraph{Imperative Primitives}

\autoref{fig:imp-prim} gives the type signatures for the imperative primitives. These are split into two groups. The first group includes the standard Idealised Algol combinators that turn \DPIA into an imperative programming language: $(\mathord;)$ sequences commands, $\prim{skip}$ is the command that does nothing; $\prim{new}~\delta$ allocates a new mutable variable on the stack, where a variable is a pair of an acceptor and an expression; and $(:=)$ assigns the value of an expression to an acceptor. For-loops are constructed by the combinators $\prim{for}$ and $\prim{parfor}$. Sequential for-loops $\prim{for}~n~b$ take a number of iterations $n$ and a loop body $b$, a command parameterised by the iteration number. Parallel for-loops $\prim{parfor}~n~\delta~a~b$ take an additional acceptor argument $a : \tyacc[n.\delta]$ that is used for the output of each iteration. The loop body for parallel for loops is required to be passive. This ensures that the side-effects of the loop body are restricted to their allotted place in the output array. This is illustrated by the non-typability of a phrase such as:
\begin{displaymath}
  \prim{parfor}~n~\delta~a~(\lambda i~o.~b := \mathsf{idx}~n~\mathsf{num}~E~i)
\end{displaymath}
where $b$ is some identifier of type $\tyacc[\mathsf{num}]$. If this loop were executed in parallel, then it would contain a data race as each parallel iteration attempted to write to $b$. Thus, by ensuring its body is passive and explicitly passing in an acceptor $o$, $\prim{parfor}$ enables deterministic data race free parallelism in an imperative setting, a key feature of \DPIA. We will see how the acceptor-transforming behaviour of our $\prim{parfor}$ primitive is translated into a normal, potentially racy, parallel for loop in \autoref{sec:translation-iii}.

Formally, newly allocated variables are zero initialised (and pointwise zero initialised for compound data), but in our implementation we typically optimise away the initialisation. In particular, it is never necessary to initialise dynamic memory allocations that are introduced by the translation of the functional primitives into imperative code as all dynamically allocated memory is always written to before being read.

The second group of imperative primitives include the acceptor variants of the $\prim{split}$, $\prim{join}$ and $\prim{pair}$ functional primitives, and array indexing. The acceptor primitives transform acceptors of compound data into acceptors of their components. They will be used to funnel data into the correct positions in the imperative translations of functional programs. In the final translation to parallel C code, described in \autoref{sec:translation-iii}, all acceptor phrases will be translated into l-values with explicit index computations.

\paragraph{Intermediate Imperative Combinators}
\autoref{fig:imp-intermediate} gives the type signatures for the intermediate imperative counterparts of $\prim{map}$ and $\prim{reduce}$. These combinators will be used in our translation from higher-order functional programs to higher-order imperative programs in \autoref{sec:translation-i}. In the second stage of the translation they will be substituted by implementations in terms of variable allocation and for-loops (\autoref{sec:translation-ii}).

\section{From Functional to Imperative}
\label{sec:translation}

As we sketched in \autoref{sec:motivation}, the translation of higher-order functional array programs to parallel C-like code happens in three stages, which we describe in this section. First, the higher-order functional combinators $\prim{map}$ and $\prim{reduce}$ are translated into the higher-order imperative combinators $\prim{mapI}$ and $\prim{reduceI}$ using an acceptor-passing translation defined mutually recursively with a continuation-passing translation (\autoref{sec:translation-i}). Secondly, the higher-order imperative combinators are translated into for-loops by substitution (\autoref{sec:translation-ii}). Finally, low-level parallel pseudo-C code is produced by translating expression and acceptor phrases into indexing operations (\autoref{sec:translation-iii}). We prove the correctness of the first two stages of our translation in \autoref{sec:correctness}.


\subsection{Translation Stage I: Higher-order Functional to Higher-order Imperative}
\label{sec:translation-i}

\begin{figure*}
  \begin{minipage}{1.0\linewidth}
    \begin{displaymath}
      \begin{array}{lcl}
        \semA{x}_\delta(A)
        &=& A :=_\delta x \medskip\\
        \semA{\underline{n}}_{\mathsf{num}}(A)
        &=& A := \underline{n} \\
        \semA{\prim{negate}~E}_{\mathsf{num}}(A)
        &=& \semE{E}_{\mathsf{num}}(\lambda x.~A := \prim{negate}~x) \\
        \semA{E_1 + E_2}_{\mathsf{num}}(A)
        &=& \semE{E_1}_{\mathsf{num}}(\lambda x.~\semE{E_2}_{\mathsf{num}}(\lambda y.~A := x + y))\medskip\\

        \semA{\prim{map}~n~\delta_1~\delta_2~F~E}_{n.\delta_2}(A)
        &=& \semE{E}_{n.\delta_1}(\lambda x.~\prim{mapI}~n~\delta_1~\delta_2~(\lambda x~o. \semA{F~x}_{\delta_2}(o))~x~A) \\
        \semA{\prim{reduce}~n~\delta_1~\delta_2~F~I~E}_{\delta_2}(A)
        &=& \begin{array}[t]{@{}l}
              \semE{E}_{n.\delta_1}(\lambda x.~\semE{I}_{\delta_2}(\lambda y.\\
              \quad\quad \prim{reduceI}~n~\delta_1~\delta_2~(\lambda x~y~o. \semA{F~x~y}_{\delta_2}(o))~y~x~(\lambda r. \semA{r}(A))))
            \end{array} \medskip\\

        \semA{\prim{zip}~n~\delta_1~\delta_2~E_1~E_2}_{n.\delta_1\times\delta_2}(A)
        &=& \semA{E_1}_{n.\delta_1}(\prim{zipAcc}_1~n~\delta_1~\delta_2~A);
            \semA{E_2}_{n.\delta_2}(\prim{zipAcc}_2~n~\delta_1~\delta_2~A)
        \\
        \semA{\prim{split}~n~m~\delta~E}_{n.m.\delta}(A)
        &=& \semA{E}_{nm.\delta}(\prim{splitAcc}~n~m~\delta~A)
        \\
        \semA{\prim{join}~n~m~\delta~E}_{nm.\delta}(A)
        &=& \semA{E}_{n.m.\delta}(\prim{joinAcc}~n~m~\delta~A)
        \\
        \semA{\prim{pair}~\delta_1~\delta_2~E_1~E_2}_{\delta_1\times\delta_2}(A)
        &=& \semA{E_1}_{\delta_1}(\prim{pairAcc_1}~\delta_1~\delta_2~A);
            \semA{E_2}_{\delta_2}(\prim{pairAcc_2}~\delta_1~\delta_2~A)
        \\
        \semA{\prim{fst}~{\delta_1}~{\delta_2}~E}_{\delta_1}(A)
        &=& \semE{E}_{\delta_1\times\delta_2}(\lambda x.~A :=_{\delta_1} \prim{fst}~{\delta_1}~{\delta_2}~x)
        \\
        \semA{\prim{snd}~{\delta_1}~{\delta_2}~E}_{\delta_2}(A)
        &=& \semE{E}_{\delta_1\times\delta_2}(\lambda x.~A :=_{\delta_2} \prim{snd}~{\delta_1}~{\delta_2}~x)
      \end{array}
    \end{displaymath}
    \subcaption{Acceptor-passing Translation}\label{fig:transone}
  \end{minipage}
  \begin{minipage}{1.0\linewidth}
    \begin{displaymath}
      \begin{array}{lcl}
        \semE{x}_\delta(C)
        &=& C(x)
        \medskip\\
        \semE{\underline{n}}_{\mathsf{num}}(C)
        &=& C(\underline{n})
        \\
        \semE{\prim{negate}~E}_{\mathsf{num}}(C)
        &=& \semE{E}_{\mathsf{num}}(\lambda x.~C(\prim{negate}~x))
        \\
        \semE{E_1 + E_2}_{\mathsf{num}}(C)
        &=& \semE{E_1}_{\mathsf{num}}(\lambda x.~\semE{E_2}_{\mathsf{num}}(\lambda y.~C(x+y)~)~)
        \medskip\\
        \semE{\prim{map}~n~\delta_1~\delta_2~F~E}_{n.\delta_2}(C)
        &=& \prim{new}~(n.\delta_2)~(\lambda \mathit{tmp}.~\semA{\prim{map}~n~\delta_1~\delta_2~F~E}_{n.\delta_2}(\mathit{tmp}.2);~C(\mathit{tmp}.1)~)
        \\
        \semE{\prim{reduce}~n~\delta_1~\delta_2~F~I~E}_{\delta_2}(C)
        &=& \begin{array}[t]{@{}l}
              \semE{E}_{n.\delta_1}(\lambda x.~
              \semE{I}_{\delta_2}(\lambda y.~\prim{reduceI}~n~\delta_1~\delta_2~(\lambda x~y~o.~\semA{F~x~y}_{\delta_2}(o))~y~x)~C)
            \end{array}
        \medskip\\
        \semE{\prim{zip}~n~\delta_1~\delta_2~E_1~E_2}_{n.\delta_1 \times \delta_2}(C)
        &=& \semE{E_1}_{n.\delta_1}(\lambda x.~\semE{E_2}_{n.\delta_2}(\lambda y.~C(\prim{zip}~n~\delta_1~\delta_2~x~y)~)~)
        \\
        \semE{\prim{split}~n~m~\delta~E}_{n.m.\delta}(C)
        &=& \semE{E}_{nm.\delta}(\lambda x.~C(\prim{split}~n~m~\delta~x)~)
        \\
        \semE{\prim{join}~n~m~\delta~E}_{nm.\delta}(C)
        &=& \semE{E}_{n.m.\delta}(\lambda x.~C(\prim{join}~n~m~\delta~x)~)
        \\
        \semE{\prim{pair}~\delta_1~\delta_2~E_1~E_2}_{\delta_1\times\delta_2}(C)
        &=& \semE{E_1}_{\delta_1}(\lambda x.~\semE{E_2}_{\delta_2}(\lambda y.~C(\prim{pair}~\delta_1~\delta_2~x~y)~)~)
        \\
        \semE{\prim{fst}~\delta_1~\delta_2~E}_{\delta_1}(C)
        &=& \semE{E}_{\delta_1\times\delta_2}(\lambda x.~C(\prim{fst}~\delta_1~\delta_2~x)~)
        \\
        \semE{\prim{snd}~\delta_1~\delta_2~E}_{\delta_2}(C)
        &=& \semE{E}_{\delta_1\times\delta_2}(\lambda x.~C(\prim{snd}~\delta_1~\delta_2~x)~)
        \\
      \end{array}
    \end{displaymath}
    \subcaption{Continuation-passing Translation}\label{fig:transtwo}
  \end{minipage}
  \caption{Acceptor and Continuation-passing Translations}
\end{figure*}

The goal of the first stage of the translation is to take a phrase $E : \mathsf{exp}[\delta]$, constructed from the functional primitives in \autoref{fig:func-prim}, and an acceptor $\mathit{out} : \mathsf{acc}[\delta]$ and to produce a $\mathsf{comm}$ phrase that has the same semantics as the command
\begin{displaymath}
  \mathit{out} :=_\delta E
\end{displaymath}
where $(:=_\delta)$ is an assignment operator for non-base types defined by induction on $\delta$ below. The resulting program will be an imperative program that acts as if we could compute the functional expression in one go and assign it to the output acceptor. Since our compilation targets know nothing of higher-order functional combinators like $\prim{map}$ and $\prim{reduce}$ they will have to be translated away. We do not use any of the traditional methods for compiling higher-order functions, such as closure conversion \cite{steele78rabbit} or defunctionalisation \cite{DBLP:journals/lisp/Reynolds98a}. Instead, we rely on the whole-program nature of our translation, our lack of recursion, and the special form of our functional primitives. Specifically, we are relying on a version of Gentzen's subformula principle (identified by \citet{Gentzen-1935} and named by \citet{Prawitz-1965}).
Our approach is reminiscent of that of \citet{NajdLSW16} who use quotation and normalisation, making essential use of the subformula principle, to embed domain-specific languages. An obvious difference with our work is that rather than stratifying a language into a host functional language and a quoted functional language, we seamlessly combine a functional and an imperative language.

We have already mentioned the use of assignment at compound data types. This is defined by:
\begin{displaymath}
  \begin{array}{lcl}
    A :=_{\mathsf{num}} E
    &=& A := E \\
    A :=_{n.\delta} E
    &=& \prim{mapI}~n~\delta~\delta~(\lambda x~a.a :=_\delta x)~E~A \\
    A :=_{\delta_1 \times \delta_2} E
    &=& \prim{pairAcc_1}~A :=_{\delta_1} \prim{fst}~E;
        \prim{pairAcc_2}~A :=_{\delta_2} \prim{snd}~E
  \end{array}
\end{displaymath}

The translation of functional expressions to imperative code is accomplished by two type-directed mutually defined translations: the acceptor-passing translation $\semA{-}_\delta$ (\autoref{fig:transone}) and the continuation-passing translation $\semE{-}_\delta$ (\autoref{fig:transtwo}). The acceptor-passing translation takes a data type $\delta$, an expression of type $\mathsf{exp}[\delta]$ and an acceptor of type $\mathsf{acc}[\delta]$, and produces a $\mathsf{comm}$ phrase. Likewise, the continuation-passing translation takes a data type $\delta$, an expression of type $\mathsf{exp}[\delta]$ and a continuation of type $\mathsf{exp}[\delta] \to \mathsf{comm}$, and produces a $\mathsf{comm}$ phrase.

It is straightforward to see by inspection, and using the fact that weakening is admissible in \DPIA, that the two translations are type-preserving:
\begin{theorem}~
\begin{enumerate}
\item If $\typd{\Pi}{\Gamma_1}{E}{\mathsf{exp}[\delta]}$ and $\typd{\Pi}{\Gamma_2}{A}{\mathsf{acc}[\delta]}$ then $\typd{\Pi}{\Gamma_1,\Gamma_2}{\semA{E}_\delta(A)}{\mathsf{comm}}$.
\item If $\typd{\Pi}{\Gamma_1}{E}{\mathsf{exp}[\delta]}$ and $\typd{\Pi}{\Gamma_2}{C}{\mathsf{exp}[\delta] \to \mathsf{comm}}$ then $\typd{\Pi}{\Gamma_1,\Gamma_2}{\semE{E}_\delta(C)}{\mathsf{comm}}$.
\end{enumerate}
\end{theorem}

It is also important that these translations satisfy the following equivalences.
{\small\begin{mathpar}
  \semA{E}_\delta(A) \simeq A :=_\delta E

  \semE{E}_\delta(C) \simeq C(E)
\end{mathpar}}%
We define observational equivalence ($\simeq$) for \DPIA and establish these particular equivalences in \autoref{sec:correctness}.
Ultimately, our goal is to compute the result of $\semA{E}_\delta(\mathit{out})$.

It might appear that we could dispense with the acceptor-passing translation and simply use $\semE{E}_\delta(\lambda x.~\mathit{out} :=_\delta x)$. However, this would create unnecessary temporary storage, violating our desire for an efficient translation. There are clear similarities between our mutually-defined translations and tail-recursive one-pass CPS translations that do not produce unnecessary administrative redexes~\cite{DanvyMN07}.

The clauses for both translations split into four groups. The first group consists only of the clause for translating functional expression phrases that are just identifiers $x$. In the acceptor-passing case, we defer to the generalised assignment defined above; in the continuation-passing case, we simply apply the continuation to the variable. The second group handles the first-order operations on numeric data. In all cases, we defer to the continuation-passing translation for the sub-expressions, with appropriate continuations.

The third group is the most interesting: the translations of the higher order $\prim{map}$ and $\prim{reduce}$ primitives. For $\prim{map}$: in the acceptor-passing case, we can immediately translate to $\prim{mapI}$ which already takes an acceptor to place its output into; in the continuation-passing case, we must create a temporary array as storage, invoke $\prim{mapI}$, and then let the continuation read from the temporary array. This indirection is required because we do not know what random access strategy the continuation $C$ will use to read the array it is given. For $\prim{reduce}$, in both cases we translate to the $\prim{reduceI}$ combinator.


The fourth group of clauses handles the translations of the functional data layout combinators. In the continuation-passing translation, they are passed straight through. They will be handled by the final translation to low-level C-like code in \autoref{sec:translation-iii}. In the acceptor-passing translation, the combinators that construct data are translated into the corresponding acceptors. In the $\prim{fst}$ and $\prim{snd}$ cases, which project out data, we defer to the continuation-passing translation. In practice, the case of a projection in tail position rarely arises, since it corresponds to disposal of part of the overall computation.






\subsection{Translation Stage II: Higher-order Imperative to For-loops}
\label{sec:translation-ii}

\newcommand{\ei}{\prim{idx}~n~(n.\delta_1)~i~E}

The next stage in the translation replaces the intermediate level imperative combinators $\prim{mapI}$ and $\prim{reduceI}$ with lower-level implementations in terms of (parallel) for-loops. This is accomplished by substitution and $\beta$-reduction (\DPIA includes full $\beta\eta$ reasoning principles). The combinator $\prim{mapI}$ is implemented as a parallel loop:
\begin{displaymath}
  \prim{mapI}
  =
  \Lambda n~\delta_1~\delta_2.\lambda F~E~A.~
  \prim{parfor}~n~\delta_2~A~(\lambda i~a.~F~(\ei)~a)
\end{displaymath}
The implementation of $\prim{reduceI}$ is more complex, involving the allocation of a temporary variable to store the accumulated value during the reduction. In this case, the for loop is sequential, since the semantics of reduction demands that we visit each element in turn:
\begin{displaymath}
  \prim{reduceI}
  =
  \Lambda n~\delta_1~\delta_2.\lambda~F~I~E~C.~\prim{new}~\delta_2~(\lambda~acc.~
    \begin{array}[t]{@{}l}
    \projtwo{acc} :=_{\delta_2} I;\\
    \prim{for}~n~(\lambda~i.~F~(\ei)~(\projone{acc})~(\projtwo{acc})\ );\\
    C~(\projone{acc})\ )
    \end{array}
\end{displaymath}
These definitions define the intended semantics of the intermediate-level imperative combinators.


\subsection{Translation Stage III: For-loops to Parallel Pseudo-C}
\label{sec:translation-iii}

\newcommand{\codegenComm}[1]{\textsc{CodeGen}_{\mathsf{comm}}(#1)}
\newcommand{\codegenAcc}[2]{\textsc{CodeGen}_{\mathsf{acc}[#1]}(#2)}
\newcommand{\codegenExp}[2]{\textsc{CodeGen}_{\mathsf{exp}[#1]}(#2)}
\newcommand{\codegenData}[1]{\textsc{CodeGen}_{\mathsf{\datatype}}(#1)}

\begin{figure}[t]
  \begin{minipage}{1.0\linewidth}
    \begin{displaymath}
      \begin{array}{l@{\hspace{0.4em}}c@{\hspace{0.4em}}l}
        \codegenComm{\prim{skip}, \eta}
        &=& \texttt{/* skip */}
        \\
        \codegenComm{\mathsf{P_1; P_2}, \eta}
        &=& \begin{array}[t]{@{}l}
              \codegenComm{P_1, \eta}~\codegenComm{P_2, \eta}
            \end{array}
        \\
        \codegenComm{A := E, \eta}
        &=& \codegenAcc{\mathsf{num}}{A, \eta, []}\texttt{ = }\codegenExp{\mathsf{num}}{E, \eta, []}\texttt{;}
        \\
        \codegenComm{\prim{new}~\delta~(\lambda v.~P), \eta}
        &=& \texttt{\{}~
            \begin{array}[t]{@{}l}
              \codegenData{\delta}~\texttt{v;} \\
              \codegenComm{P[(v_a,v_e)/v], \eta[v_a \mapsto \texttt{v}, v_e \mapsto \texttt{v}]}\ \texttt{\}}
            \end{array}
        \\
        \codegenComm{\prim{for}~n~(\lambda i.~P), \eta}
        &=& \begin{array}[t]{@{}l}
              \texttt{for(int i = 0; i < $n$; i += 1) \{} \\
              \quad \codegenComm{P, \eta[i \mapsto \texttt{i}]} \ \texttt{\}}
            \end{array}
        \\
        \codegenComm{\prim{parfor}~n~\delta~A~(\lambda i~o.~P)~E, \eta}
        &=& \begin{array}[t]{@{}l}
              \texttt{parfor(int i = 0; i < $n$; i += 1) \{ }\\
              \quad \codegenComm{P[\prim{idxAcc}~n~\delta~A~i/o], \eta[i \mapsto \texttt{i}]}\ \texttt{\}}
            \end{array}
      \end{array}
    \end{displaymath}
    \subcaption{\DPIA commands to C statements}\label{fig:codegen-comm}
  \end{minipage}

  \bigskip

  \begin{minipage}{1.0\linewidth}
    \begin{displaymath}
      \begin{array}{@{}l@{~}c@{~}l@{}}
        \codegenAcc{\delta}{x, \eta, \mathit{ps}}
        &=& \eta(x)(\mathit{reverse}~\mathit{ps})
        \\
        \codegenAcc{\mathsf{\delta}}{\prim{idxAcc}~n~\delta~A~I, \eta, \mathit{ps}}
        &=& \begin{array}[t]{@{}l}
          \codegenAcc{\mathsf{n.\delta}}{A, \eta,\\ \quad\codegenExp{\mathit{idx}(n)}{I, \eta, []} :: \mathit{ps})}
        \end{array}
        \\
        \codegenAcc{\mathsf{nm.\delta}}{\prim{splitAcc}~n~m~\delta~A, \eta, \texttt{i} :: \mathit{ps}}
        &=& \codegenAcc{\mathsf{n.m.\delta}}{A, \eta, \texttt{i/}m :: \texttt{i\%}m :: \mathit{ps}}
        \\
        \codegenAcc{\mathsf{n.m.\delta}}{\prim{joinAcc}~n~m~\delta~A, \eta, \texttt{i} :: \texttt{j} :: \mathit{ps}}
        &=& \codegenAcc{\mathsf{nm.\delta}}{A, \eta, \texttt{i*}m\texttt{+j} :: \mathit{ps}}
        \\
        \codegenAcc{\delta_1}{\prim{pairAcc_1}~\delta_1~\delta_2~A, \eta, \mathit{ps}}
        &=& \codegenAcc{\delta_1\times\delta_2}{A, \eta, \texttt{.x1} :: \mathit{ps}}
        \\
        \codegenAcc{\delta_2}{\prim{pairAcc_2}~\delta_1~\delta_2~A, \eta, \mathit{ps}}
        &=& \codegenAcc{\delta_1\times\delta_2}{A, \eta, \texttt{.x2} :: \mathit{ps}}
        \\
        \codegenAcc{n.\delta_1}{\prim{zipAcc_1}~n~\delta_1~\delta_2~A, \eta, \texttt{i} :: \mathit{ps}}
        &=& \codegenAcc{n.(\delta_1\times\delta_2)}{A, \eta, \texttt{i} :: \texttt{.x1} :: \mathit{ps}}
        \\
        \codegenAcc{n.\delta_2}{\prim{zipAcc_2}~n~\delta_1~\delta_2~A, \eta, \texttt{i} :: \mathit{ps}}
        &=& \codegenAcc{n.(\delta_1\times\delta_2)}{A, \eta, \texttt{i} :: \texttt{.x2} :: \mathit{ps}}
      \end{array}
    \end{displaymath}
    \subcaption{\DPIA acceptors to C l-values}\label{fig:codegen-acc}
  \end{minipage}

  \bigskip

  \begin{minipage}{1.0\linewidth}
    \begin{displaymath}
      \begin{array}{@{}l@{~}c@{~}l@{}}
        \codegenExp{\delta}{x, \eta, \mathit{ps}}
        &=& \eta(x)(\mathit{reverse}~\mathit{ps})
        \\
        \codegenExp{\mathsf{num}}{\underline{n}, \eta, []}
        &=& n
        \\
        \codegenExp{\mathsf{num}}{\prim{negate}~E, \eta, []}
        &=& \texttt{(}\texttt{-}~\codegenExp{\mathsf{num}}{E, \eta, []}\texttt{)}
        \\
        \codegenExp{\mathsf{num}}{E_1 + E_2, \eta, []}
        &=& \texttt{(}
        \begin{array}[t]{@{}l}
          \codegenExp{\mathsf{num}}{E_1, \eta, []}\\
          \texttt{+ }\codegenExp{\mathsf{num}}{E_2, \eta, []}\texttt{)}
        \end{array}
        \\
        \codegenExp{n.(\delta_1 \times \delta_2)}{\prim{zip}~n~\delta_1~\delta_2~E_1~E_2, \eta, \texttt{i} :: \texttt{.x}j :: \mathit{ps}}
        &=& \codegenExp{n.\delta_j}{E_j, \eta, \texttt{i} :: \mathit{ps}}
        \\
        \codegenExp{m.n.\delta}{\prim{split}~n~m~\delta~E, \eta, \texttt{i} :: \texttt{j} :: \mathit{ps}}
        &=& \codegenExp{mn.\delta}{E, \eta, \texttt{i*}n\texttt{+j} :: \mathit{ps}}
        \\
        \codegenExp{mn.\delta}{\prim{join}~n~m~\delta~E, \eta, \texttt{i} :: \mathit{ps}}
        &=& \codegenExp{m.n.\delta}{E, \eta, \texttt{i/}n :: \texttt{i\%}n :: \mathit{ps}}
        \\
        \codegenExp{\delta_1 \times \delta_2}{\prim{pair}~\delta_1~\delta_2~E_1~E_2, \eta, \texttt{.x}j :: \mathit{ps}}
        &=& \codegenExp{\delta_j}{E_j, \eta, \mathit{ps}}
        \\
        \codegenExp{\delta_1}{\prim{fst}~\delta_1~\delta_2~E, \eta, \mathit{ps}}
        &=& \codegenExp{\delta_1 \times \delta_2}{E, \eta, \texttt{.x1} :: \mathit{ps}}
        \\
        \codegenExp{\delta_2}{\prim{snd}~\delta_1~\delta_2~E, \eta, \mathit{ps}}
        &=& \codegenExp{\delta_1 \times \delta_2}{E, \eta, \texttt{.x2} :: \mathit{ps}}
        \\
        \codegenExp{\delta}{\prim{idx}~n~\delta~E~I, \eta, \mathit{ps}}
        &=&
        \begin{array}[t]{@{}l}
          \textsc{CodeGen}_{\mathsf{exp}[n.\delta]}(E, \eta,\\ \quad\codegenExp{\mathit{idx}(n)}{I, \eta, []} :: \mathit{ps})
        \end{array}
      \end{array}
    \end{displaymath}
    \subcaption{\DPIA expressions to C r-values}\label{fig:codegen-exp}
  \end{minipage}

  \caption{Translation of Purely Imperative \DPIA to Parallel Pseudo C}
  \label{fig:codegen}
\end{figure}

After performing the translation steps in the previous two sections, we have generated a command phrase that does not use the higher-order functional combinators $\prim{map}$ and $\prim{reduce}$, but still contains uses of the data layout combinators $\prim{zip}$, $\prim{split}$ etc., and their acceptor counterparts. We now define a translation to a C-like language with parallel for loops that resolves these data layout expressions into explicit indexing expressions.  The translation is defined in \autoref{fig:codegen}. Parallel for loops can easily be achieved using OpenMP's \texttt{\#pragma parallel for} construct, for example. In \autoref{sec:opencl}, we describe how to adapt this process (and the earlier stages) to work with the real parallel C-like language OpenCL.

The translation in \autoref{fig:codegen} is split into three parts: the translation of \DPIA commands into C statements, the translation of acceptors into l-values, and the translation of expressions into r-values. (Recall the analogy made in \autoref{sec:types} between the phrase types of \DPIA and syntactic categories in an imperative language; we are now reaping the rewards of Reynolds' careful design of Idealised Algol.) We assume that the input to the translation process is in $\beta\eta$-normal form, so all $\prim{new}$-block and loop bodies are fully expanded.

The translation of commands in \autoref{fig:codegen-comm} straightforwardly translates each \DPIA command into the corresponding C statement. The translation is parameterised by an environment $\eta$ that maps from \DPIA identifiers to C variable names. There is a small discrepancy that we overcome in that semicolons are statement terminators in C, but separators in \DPIA, and that doing nothing useful is unremarkable in a C program, but is explicitly written $\prim{skip}$ in \DPIA. The translation of assignment relies on the translations for acceptors and expression, which we define below. Variable allocation is translated using a \texttt{\{ ... \}} block to limit the scope. We omit initialisation of the new variable because we know by inspection of our previous translation steps that newly allocated variables will always be completely initialised before reading. Note how we explicitly substitute a pair of identifiers for the acceptor and expression parts of the variable in \DPIA, but that these both refer to the same C variable in the extended environment. Translation of variable allocation makes use of generator $\codegenData{\delta,\texttt{v}}$ that generates the appropriate C-style variable declaration for the data type $\delta$. Since C does not have anonymous tuple types, this entails on-the-fly generation of the appropriate \texttt{struct} definitions.

\DPIA $\prim{for}$ loops are translated into C \texttt{for} loops, \DPIA $\prim{parfor}$ loops are translated into pseudo-parallel-for loops. In the body of the \texttt{parfor} loop, we substitute in an $\prim{idxAcc}$ phrase which will be resolved later by the translation of acceptors.

The variable names introduced by translating $\prim{new}$, $\prim{for}$ and $\prim{parfor}$ are all assumed to be fresh.

The translation of acceptors (\autoref{fig:codegen-acc}) is parameterised by an environment $\eta$, as for commands, as well as a path $\mathit{ps}$, consisting of a list of C-expressions of type $\texttt{int}$ denoting array indexes, and \texttt{struct} fields, \texttt{.x1}, \texttt{.x2}, denoting projections from pairs. The path must always agree with the type of the acceptor being translated. During command translation, all acceptors being translated have type $\texttt{num}$ so the access path starts empty. The acceptor translation clauses in \autoref{fig:codegen-acc} all proceed by manipulating the access path appropriately until an identifier is reached. At this point, the \DPIA identifier is replaced with its corresponding C variable and the access path is appended.

Translation of expressions (\autoref{fig:codegen-exp}) is parameterised similarly to the acceptor translation, and contains similar clauses for all the data layout combinators. Expressions also include literals and arithmetic expressions, which are translated to the corresponding notion in C.

\newcommand{\etaext}{\eta[i \mapsto \texttt{i}]}

\paragraph{Example} We demonstrate how the translation to C works by applying it to the $\prim{parfor}$ loop in the translation (\ref{codeex:dpia-dot-product}) of the simple dot product in \autoref{sec:motivation}. We use the environment $\eta = [\mathit{out} \mapsto \texttt{out}, \mathit{xs} \mapsto \texttt{xs}, \mathit{ys} \mapsto \texttt{ys}]$. The command translation translates, in two steps, the $\prim{parfor}$ loop and the assignment, substituting in the indexing acceptor for the acceptor in the loop body:
\begin{displaymath}
  \begin{array}{cl}
    &\codegenComm{\prim{parfor}~\underline{n}~\mathit{out}~(\lambda i~o.
      o := \prim{fst}~(\prim{idx}~(\prim{zip}~\mathit{xs}~\mathit{ys})~i) *
      \prim{snd}~(\prim{idx}~(\prim{zip}~\mathit{xs}~\mathit{ys})~i)), \eta}
    \\
    =& \begin{array}[t]{@{}l}
         \texttt{parfor(int i = 0; i < }n\texttt{; i+=1) \{}\\
         \quad \codegenComm{\prim{idxAcc}~\mathit{out}~i := \prim{fst}~(\prim{idx}~(\prim{zip}~\mathit{xs}~\mathit{ys})~i) *
      \prim{snd}~(\prim{idx}~(\prim{zip}~\mathit{xs}~\mathit{ys})~i)), \etaext}\\
         \texttt{\}}
       \end{array}
    \\
    =& \begin{array}[t]{@{}l}
         \texttt{parfor(int i = 0; i < }n\texttt{; i+=1) \{}\\
         \quad \codegenAcc{\mathsf{num}}{\prim{idxAcc}~\mathit{out}~i, \etaext, []} =\\
         \qquad \codegenExp{\mathsf{num}}{\prim{fst}~(\mathsf{idx}~(\prim{zip}~\mathit{xs}~\mathit{ys})~i) *
         \prim{snd}~(\prim{idx}~(\prim{zip}~\mathit{xs}~\mathit{ys})~i)), \etaext, []}\texttt{;}\\
         \texttt{\}}
       \end{array}
  \end{array}
\end{displaymath}
The acceptor part of the assignment is translated as follows:
\begin{displaymath}
  \begin{array}{lcl}
    \codegenAcc{\mathsf{num}}{\prim{idxAcc}~\mathit{out}~i, \etaext, []}
    &=&\codegenAcc{\underline{n}.\mathsf{num}}{\mathit{out}, \etaext, [\texttt{i}]} \\
    &=&\texttt{out[i]}
  \end{array}
\end{displaymath}
The expression part of the assignment is translated as follows, where we only spell out the left-hand side of the multiplication in detail; the right hand side is similar.
\begin{displaymath}
  \begin{array}{cl}
    &\codegenExp{\mathsf{num}}{\prim{fst}~(\prim{idx}~(\prim{zip}~\mathit{xs}~\mathit{ys})~i), \etaext, []} \\
    =&\codegenExp{\mathsf{num}\times\mathsf{num}}{\prim{idx}~(\prim{zip}~\mathit{xs}~\mathit{ys})~i, \etaext, [\texttt{.x1}]} \\
    =&\codegenExp{\underline{n}.(\mathsf{num}\times\mathsf{num})}{\prim{zip}~\mathit{xs}~\mathit{ys}, \etaext, [\texttt{i}, \texttt{.x1}]} \\
    =&\codegenExp{\underline{n}.\mathsf{num}}{\mathit{xs}, \etaext, [\texttt{i}]} \\
    =&\texttt{xs[i]}
  \end{array}
\end{displaymath}
Putting everything together, we get the following translation of the original $\prim{parfor}$ loop, which has had all the data layout combinators translated away.
\begin{displaymath}
    \texttt{parfor(int i = 0; i < $n$; i+=1) \{ out[i] = xs[i] * ys[i]; \}}
\end{displaymath}

%
%

A similar translation was recently presented in a more informal style by \citet{SteuwerReDu2017/cgo}.
Their experimental results show that in practice it is important to keep the indices concise and short for generating efficient OpenCL code and discuss how to simplify index expressions making use of range information of the indices involved.





\section{Correctness of the Translation}
\label{sec:correctness}

We now justify the translation process described in \autoref{sec:translation-i} and \autoref{sec:translation-ii}. We have not yet formalised a correctness proof for the final translation to C (\autoref{sec:translation-iii}); this is future work.

As we stated in \autoref{sec:translation}, the goal of the translation process was to generate a ``purely imperative'' command $\semA{E}_\delta(A)$ that is equivalent to the assignment command $A :=_\delta E$. To make this statement formal, we must define equivalence of \DPIA programs. We do this in \autoref{sec:equivalence}. We then state the functional coincidence property that establishes that our correctness property means just what we intend (\autoref{sec:coincidence}). Finally, we prove the correctness of our translation in \autoref{sec:correctness-proof}.

\subsection{Semantics and Observational Equivalence for \DPIA}
\label{sec:equivalence}

We use Reddy's relationally parametric automata-theoretic semantics of SCIR \cite{DBLP:journals/entcs/Reddy13}. In this model, the interpretations of phrase types are parameterised by automata describing the permitted state transitions. The model supports relationally parametric reasoning \cite{reynolds83types}, which enables reasoning about locality and information hiding (following \cite{DBLP:journals/jacm/OHearnT95}) and the use of automata permits reasoning about phrases that do not affect the state (i.e. passive phrases).


Reddy's model does not include indexed types or compound data types, as we do in \DPIA, but these are straightforward to add to the model. Indexed types are interpreted by set-theoretic functions whose codomain depends on the input (we do not interpret the data type polymorphism using parametricity because we are only interested in parametric reasoning for local state). Compound data types are interpreted as the following sets when they appear in expressions:
{\small
\begin{mathpar}
  \sem{\mathsf{num}} = \mathrm{num}

  \sem{\delta_1 \times \delta_2} = \sem{\delta_1} \times \sem{\delta_2}

  \sem{\underline{n}.\delta} = \{0,\dots,n-1\} \to \sem{\delta}
\end{mathpar}}
where the set $\mathrm{num}$ is some set of number-like objects used for scalar values. This interpretation of data types allows us to straightforwardly interpret the functional primitives of \autoref{fig:func-prim} in the model, using the standard interpretation of $\prim{map}$ and $\prim{reduce}$ explicitly given in \cite{SteuwerFeLiDu2015/icfp}. This yields a coincidence property that we state in \autoref{sec:coincidence} below. Acceptors for compound data types are interpreted using the separating product construction, which ensures that disjoint components of compound acceptors are always non-interfering. This allows us to interpret $\prim{parfor}~n$ as $n$ parallel transformations on $n$ pieces of disjoint state.

Reddy's semantics assigns an interpretation to every \DPIA phrase. For observational equivalence of \DPIA programs, we are interested in closed programs. A closed program is a command phrase whose free identifiers are all have types of the form $\tyvar[\delta]$ for possibly different $\delta$. Our notion of observational equivalence is standard. Two well-typed phrases $\typd{\Pi}{\Gamma}{P_1, P_2}{\theta}$ are equivalent iff for all closing contexts $C[-]$, the programs $C[P_1]$ and $C[P_2]$, when instantiated with the standard interpretation of variables (\cite{DBLP:journals/entcs/Reddy13}, Figure 2), describe the same mapping of initial to final values. We formally write $\typd{\Pi}{\Gamma}{P_1 \simeq P_2}{\theta}$, or informally $P_1 \simeq P_2$. Note that this relation is automatically an equivalence and is congruent with all the constructs of \DPIA.

For the purposes of compilation, this notion of equivalence is justifiable: we are only interested in the relationship between the initial and final values of each variable, not the intermediate states.

\subsection{Functional Coincidence}
\label{sec:coincidence}

Our correctness criterion will have no force if we do not first establish that the assignment $\mathit{out} :=_\delta E$ means what we think it means.
We state our coincidence property formally as follows.  Let $\typ{\cdot}{x_1 : \tyexp[\delta_1], ..., x_n : \tyexp[\delta_n]}{\cdot}{E}{\tyexp[\delta]}$ be some expression phrase of \DPIA built from the functional primitives in \autoref{fig:func-prim}. Let $\sem{E} : \sem{\delta_1} \times ... \times \sem{\delta_n} \to \sem{\delta}$ be the functional reference semantics of $E$. Use $E$ to construct a closed phrase:
\begin{displaymath}
  \typ{\cdot}{\cdot}{v_1 : \tyvar[\delta_1], ..., v_n : \tyvar[\delta_n], \mathit{out} : \tyvar[\delta]}{\mathit{out}.1 :=_\delta E[v_1.2/x_1, ..., v_n.2/x_n]}{\tycomm}
\end{displaymath}
Then for all $a_1 \in \sem{\delta_1}, ..., a_n \in \sem{\delta_n}, a \in \sem{\delta}$, the interpretation of this command maps the store $(\mathit{v_1} \mapsto a_1, ..., v_n \mapsto a_n, \mathit{out} \mapsto a)$ to the store $(\mathit{v_1} \mapsto a_1, ..., v_n \mapsto a_n, \mathit{out} \mapsto \sem{E}(a_1, ..., a_n))$. In other words, this program updates the variable $\mathit{out}$ with the result of the expression, and leaves every other variable unaffected. We use \citet{SteuwerFeLiDu2015/icfp}'s interpretation of the functional primitives in our \DPIA semantics, so this property is immediate.

\subsection{Correctness of the Translation from Functional to Imperative}
\label{sec:correctness-proof}

We structure our proof by first stating a collection of equivalences that can be proved in Reddy's model, and then use them to prove that the translation of \autoref{sec:translation-i} and \autoref{sec:translation-ii} is correct (\autoref{thm:correctness}). The properties of contextual equivalences that we use in our proof, in addition to the fact that $\simeq$ is a congruent equivalence relation, are as follows.
\begin{enumerate}
\item $\beta\eta$-equality for non-dependent and dependent functions:
  {\small\begin{mathpar}
    (\lambda x.~P)Q \simeq P[Q/x]

    P \simeq (\lambda x.~P x)

    (\Lambda x.~P)e \simeq P[e/x]

    P \simeq (\Lambda x.~P x)
  \end{mathpar}}
  Full $\beta\eta$-equality for functions is one of the defining features of Idealised Algol and its descendents \cite{Reynolds97}. These are all justified by Reddy's model (and indeed almost all models of Idealised Algol-like languages).
\item The $\prim{parfor}$-based implementation of $\prim{mapI}$ (\autoref{sec:translation-ii}) satisfies the following equivalence:
  \begin{smallequation}\label{propty:map}
    \prim{mapI}~n~\delta_1~\delta_2~(\lambda x~o.~o:=_{\delta_2}F~x)~E~A
    \quad \simeq \quad
    A :=_{n.\delta_2} \prim{map}~n~\delta_1~\delta_2~F~E
  \end{smallequation}
  By the definition of array assignment given in \autoref{sec:translation-i}, this property is equivalent to:
  \begin{displaymath}
    \prim{mapI}~n~\delta_1~\delta_2~(\lambda x~o.~o:=_{\delta_2}F~x)~E~A
    \quad \simeq \quad
    \prim{mapI}~n~\delta_2~\delta_2~(\lambda x~o.~o:=_{\delta_2}x)~(\prim{map}~\delta_1~\delta_2~F~E)~A
  \end{displaymath}
  Expanding the definition of $\prim{mapI}$, and $\beta$-reducing, we must show:
  \begin{displaymath}
    \qquad\prim{parfor}~n~A~(\lambda i~o.~o:=_{\delta_2}F~(\prim{idx}~n~\delta_1~E~i))
    \simeq
    \prim{parfor}~n~A~(\lambda i~o.~o:=_{\delta_2}\prim{idx}~n~\delta_2~(\prim{map}~n~\delta_1~\delta_2~F~E)~i)
  \end{displaymath}
  which is immediate from the way that array data types are interpreted as functions from indices to values.

\item Reddy's model validates the following equivalence involving the use of temporary storage. For all expressions $E$ and continutations $C$ that are non-interfering, we have:
  \begin{smallequation}\label{propty:temp-storage}
    \prim{new}~\delta~(\lambda \mathit{tmp}.~\mathit{tmp}.1 :=_\delta E; C(\mathit{tmp}.2))
    \quad \simeq \quad
    C(E)
  \end{smallequation}
  This equivalence relies crucially on the fact that $C$ and $E$ cannot interfere, so we can take a complete copy of $E$ before invoking $C$. If $C$ were able to write to storage that is read by $E$, then it would not be safe to cache $E$ before invoking $C$. In Reddy's model, we use parametricity to relate the two uses of $C$: one in a store that contains the state that $E$ reads, and one in a store that contains the result of evaluating $E$. Using parametricity and restriction to only the identity state transition on $E$'s portion of the store further ensures that $C$ does not interfere with $E$.

\item The $\prim{for}$-loop based implementation of $\prim{reduceI}$ should satisfy the following equivalence.
  \begin{smallequation}\label{propty:reduce}
    \prim{reduceI}~n~\delta_1~\delta_2~(\lambda x~y~o.~o :=_{\delta_2}F~x~y)~I~E~C
    \quad\simeq\quad
    C(\prim{reduce}~n~\delta1~\delta_2~F~I~E)
  \end{smallequation}
  Substituting in the implementation of $\prim{reduceI}$, and $\beta$-reducing, this is equivalent to showing:
  \begin{displaymath}
    \prim{new}~\delta_2~(\lambda v.~v.1 :=_{\delta_2} I; \prim{for}~n~(\lambda i.~v.1 :=_{\delta_2} F~v.2~(\prim{idx}~E~i)); C(v.2))
    \simeq
    C(\prim{reduce}~n~\delta1~\delta_2~F~I~E)
  \end{displaymath}
  Because the acceptor-expression pair $v$ has been freshly allocated, it acts like a so-called ``good variable'' in Idealised Algol terminology. This means that the following equivalence holds, using the fact that neither $F$ nor $E$ interfere with $v$:
  \begin{displaymath}
    \prim{for}~n~(\lambda i.~v.1 :=_{\delta_2} F~v.2~(\prim{idx}~E~i))
    \simeq
    v.1 :=_{\delta_2} \prim{reduce}~n~\delta_1~\delta_2~F~v.2~E
  \end{displaymath}
  Now \autoref{propty:reduce} follows from \autoref{propty:temp-storage}.
\item Finally, we need agreement between the data layout combinators and their acceptor counterparts:
  \begin{displaymath}
    \begin{array}{lcl}
      A :=_{\delta_1 \times \delta_2}~\prim{pair}~\delta_1~\delta_2~E_1~E_2
      &\simeq&(\prim{pairAcc_1}~\delta_1~\delta_2~A :=_{\delta_1} E_1;
               \prim{pairAcc_2}~\delta_1~\delta_2~A :=_{\delta_2} E_2)
      \\
      A :=_{n.(\delta_1 \times \delta_2)}~\prim{zip}~n~\delta_1~\delta_2~E_1~E_2
      &\simeq& (\prim{zipAcc_1}~n~\delta_1~\delta_1~A :=_{n.\delta_1} E_1;
               \prim{zipAcc_2}~n~\delta_1~\delta_1~A :=_{n.\delta_2} E_2)
      \\
      A :=_{n.m.\delta} \prim{split}~n~m~\delta~E
      &\simeq& \prim{splitAcc}~n~m~\delta~A :=_{nm.\delta} E
      \\
      A :=_{nm.\delta} \prim{join}~n~m~\delta~E
      &\simeq& \prim{joinAcc}~n~m~\delta~A :=_{n.m.\delta} E
    \end{array}
  \end{displaymath}
  The first equivalence follows directly from the definition of assignment at pair type, and $\beta$-reduction for pairs. The others are all straightforwardly justified in Reddy's model, given the interpretation of acceptors for compound data types using separating products, described above.
\end{enumerate}

\begin{theorem}
  \label{thm:correctness}
  The translations $\semA{-}_-(-)$ and $\semE{-}_-(-)$ defined in \autoref{fig:transone} and \autoref{fig:transtwo} satisfy the following observational equivalences for all acceptors $A$ and functional expressions $E$ with disjoint sets of active identifiers:
  {\small\begin{mathpar}
    \semA{E}_\delta(A) \simeq A :=_\delta E

    \semE{E}_\delta(C) \simeq C(E)
  \end{mathpar}}
\end{theorem}

\begin{proof}
  By mutual induction on the steps of the translation process. The cases for variables in both translations are immediate. The cases for the first-order combinators on numbers follow from the induction hypotheses and $\beta$-reduction. For example, for $\prim{negate}$:
  \begin{displaymath}
    \semA{\prim{negate}~E}_{\mathsf{num}}(A)
    = \semE{E}_{\mathsf{num}}(\lambda x.~A := \prim{negate}~x)
    \simeq (\lambda x.~A := \prim{negate}~x)(E)
    \simeq A := \prim{negate}~E
  \end{displaymath}
  The case for the first-order combinators in the continuation-passing translation are similar.

  The acceptor-passing translation of $\prim{map}$ uses the induction hypothesis to establish the correctness of the translations of the subterms, $\beta$-equality, and then the correctness property of $\prim{mapI}$ (\autoref{propty:map}):
  \begin{displaymath}
    \begin{array}{lcl}
      \semA{\prim{map}~n~\delta_1~\delta_2~F~E}_{n.\delta_2}(A)
      &=&\semE{E}_{n.\delta_1}(\lambda x.~\prim{mapI}~n~\delta_1~\delta_2~(\lambda x~o.~\semA{F~x}_{\delta_2}(o))~x~A) \\
      &\simeq& \prim{mapI}~n~\delta_1~\delta_2~(\lambda x~o.~\semA{F~x}_{\delta_2}(o))~E~A) \\
      &\simeq& \prim{mapI}~n~\delta_1~\delta_2~(\lambda x~o.~o:=_{\delta_2} F~x)~E~A \\
      &\simeq& A :=_{n.\delta_2} \prim{map}~n~\delta_1~\delta_2~F~E
    \end{array}
  \end{displaymath}
  The continuation-passing translation of $\prim{map}$ relies on the acceptor-passing translation and additionally \autoref{propty:temp-storage} that using temporary storage is unobservable. The acceptor-passing and contination-passing translations for $\prim{reduce}$ both rely on \autoref{propty:reduce} establishing the correctness of $\prim{reduceI}$.

  The acceptor-passing translations of the data layout combinators rely on the corresponding properties for $\prim{zip}$, $\prim{split}$, $\prim{join}$ and $\prim{pair}$. The acceptor-passing cases for $\prim{fst}$ and $\prim{snd}$ follow from the induction hypothesis and $\beta$-equality. The correctness of the contination-passing translations for the data layout combinators also follow by applying the induction hypothesis and using $\beta$-equality.
\end{proof}



\section{From Data Parallel Idealised Algol to OpenCL}
\label{sec:opencl}

In \autoref{sec:translation} we discussed the translation of higher-order functional array programs to imperative combinators \mapI and \reduceI which are then expanded into for-loops by substitution.
In \autoref{sec:correctness} we have shown that this translation is semantics preserving.
In this section, we discuss the process of OpenCL code generation.
OpenCL~\citep{opencl} is the leading standard for programming GPUs and accelerators.
These devices offer tremendous compute power for many application domains, including mathematical finance and deep learning, which makes GPUs an important hardware target.

\subsection{A Short Introduction to OpenCL}
\label{sec:opencl:opencl}

The OpenCL programming model distinguishes between the managing \emph{host program} and the \emph{kernel programs} which are executed on parallel on an OpenCL enabled \emph{device}.
Kernel programs are special functions written in the OpenCL C programming language which is a dialect of C with parallel-specific restrictions and extensions.
Our work focuses purely on the generation of the OpenCL kernel.
A kernel function is executed in parallel on an OpenCL device by multiple \emph{work-items} (threads) which can optionally be organised in \emph{work-groups}.
Each work-item is uniquely identified by a \emph{global id}, or a combination of a \emph{group id} and a \emph{local id} internal to the group.
These ids are used to determine which part of the data is accessed by each threads.

OpenCL also defines different \emph{memory spaces} which correspond to memories with distinct performance characteristics.
The \emph{global memory} is visible by all the threads and is usually the largest, but also the slowest memory on an OpenCL device.
The \emph{local memory} is shared among the work-items of a work-group and is order of magnitudes faster than global memory (comparable to cache performance).
Finally, the \emph{private memory} is the fastest memory, but very small and can not be used for data shared among work-items (private memory usually corresponds to registers).

On some architectures vector instructions are crucial for achieving high performance.
OpenCL supports special \emph{vector types} such as \texttt{float4} where operations on a value of this type are performed by the vector units in the processor.
Vector types are only available for a small number of underlying numerical data types (\eg \texttt{int}, \texttt{float}) and a fixed number of sizes: 2, 3, 4, 8, and 16.

\subsection{OpenCL Specific Data Parallel Programming Primitives}

Following the work of \citet{SteuwerFeLiDu2015/icfp} we have designed a set of parallel programming primitives reflecting the OpenCL programming model in an extension of \DPIA.
Their work has shown that the design presented below allows generation of efficient OpenCL code with performance comparable to expert written code.

\paragraph{Parallelism Hierarchy}
To exploit the different parallelism levels of the OpenCL thread hierarchy with global work-items, local work-items organised in work-groups, and sequential execution inside a single work-item, we introduce four variants of the functional $\prim{map}$ primitive, all with the same type as the original:

\begin{displaymath}
  \quad\left.\begin{array}{@{}l@{~}}
        \prim{mapGlobal}\\ \prim{mapWorkgroup}\\ \prim{mapLocal}\\ \prim{mapSeq}
  \end{array}\right\}
  : (n : \natkind) \to (\delta_1~\delta_2 : \datatype) \to (\tyexp[\delta_1] \to \tyexp[\delta_2]) \to \tyexp[n.\delta_1] \to \tyexp[n.\delta_2]\mathrightfill
\end{displaymath}
We also add four corresponding intermediate imperative combinators, specialising the $\prim{mapI}$ used above:
\begin{displaymath}
  \quad\left.\begin{array}{@{}l@{~}}
    \prim{mapIGlobal}\\ \prim{mapIWorkgroup}\\\prim{mapILocal}\\\prim{mapISeq}
  \end{array}\right\}
  :
  \begin{array}{@{}l@{}}
    (n : \natkind) \to (\delta_1~\delta_2 : \datatype) \to
      (\tyexp[\delta_1] \to \tyacc[\delta_2] \pureto \tycomm) \to\\ \qquad \tyexp[n.\delta_1] \to \tyacc[n.\delta_2] \to \tycomm
  \end{array}\mathrightfill
\end{displaymath}
Finally, we add three OpenCL-specific variations of the $\prim{parfor}$ imperative primitive:
\begin{displaymath}
  \quad\left.\begin{array}{@{}l@{~}}
    \prim{parforGlobal}\\\prim{parforWorkgroup}\\\prim{parforLocal}
    \end{array}\right\}
  :
  (n : \natkind) \to (\delta : \datatype) \to \tyacc[n.\delta] \to (\tyexp[\mathrm{idx}(n)] \to \tyacc[\delta] \pureto \tycomm) \to \tycomm\mathrightfill
\end{displaymath}
We reuse the sequential $\prim{for}$ primitive for the translation of $\prim{mapSeq}$.
The specification of the translation of the specialised $\prim{map}*$ functional primitives down to the corresponding variations of $\prim{parfor}$ via their intermediate imperative counterparts is defined exactly as for the $\prim{map} \to \prim{mapI} \to \prim{parfor}$ translation in \autoref{sec:translation}. Semantically, all these variants of $\prim{map}$ and $\prim{parfor}$ are equivalent to the originals, so the correctness proof in \autoref{sec:correctness} is unaffected. The additional information present in the names is only used by the OpenCL code generator. However, in future work, we want to formalise the OpenCL model to ensure by construction that we always generate valid OpenCL kernels that respect the parallelism hierarchy.

\paragraph{Address Spaces}
\label{sec:opencl:addrspace}
To account for OpenCL's multiple address spaces, we add three primitives
which wrap a function, i.\,e., take a function as its argument and return a function of the same type:
\begin{displaymath}
\quad
    \prim{toGlobal},\prim{toLocal},\prim{toPrivate}
  :
  (\delta_1~\delta_2 : \datatype) \to (\tyexp[\delta_1] \to \tyexp[\delta_2]) \to \tyexp[\delta_1] \to \tyexp[\delta_2]\mathrightfill
\end{displaymath}
Semantically, these functions are all the identity. As above, the additional information is only used by the OpenCL code generator.
During the translation these functions are replaced by specialised $\prim{new}$ primitives parameterised with the OpenCL memory space and perform the memory allocation in the indicated memory space:
\begin{displaymath}
  \quad
    \prim{newGlobal},\prim{newLocal},\prim{newPrivate}
  :
  (\delta : \datatype) \to (\tyvar[\delta] \to \tycomm) \to \tycomm \mathrightfill
\end{displaymath}
By default, $\prim{map}$ allocates memory in global memory for its output during the continuation-passing translation.
When $\prim{map}$ is wrapped in, e.g., $\prim{toLocal}$ this will perform the memory allocation and trigger the acceptor-passing translation of $\prim{map}$ where it does not allocate memory itself, but rather writes to the provided acceptor.

As for the parallelism hierarchy, in future work we plan to extend our formal treatment to include the OpenCL memory model and track address space use with an effect system.
This will allow us to ensure that the address spaces are only used correctly.

\paragraph{Vectorisation}
To support the OpenCL vector types we extended \DPIA's type system with an additional vector data type.
This is defined similar to the array data type, but more restricted so that the element data type has to be $\mathsf{num}$ and the length must be one of the legal choices defined by OpenCL.
Arrays of non-vector type can be turned into an array of vector type using the $\prim{asVector}$ primitive which behaves similar to the $\prim{split}$ primitive:
\begin{displaymath}
  \quad\begin{array}{@{}l@{}}\prim{asVector}_{\underline{n}}\end{array}
  :
  (m : \natkind) \to (\delta : \datatype) \to \tyexp[m\underline{n}.\delta] \to \tyexp[m.\mathsf{num}\langle \underline{n} \rangle]\quad (\text{where } \mathsf{num}\langle \underline{n} \rangle \text{ is a vector type})\mathrightfill
\end{displaymath}
Similarly to $\prim{join}$ which flattens a two dimensional array, $\prim{asScalar}$ turns an array of vector type into an array of non-vector type:
\begin{displaymath}
  \quad\begin{array}{@{}l@{}}\prim{asScalar}_{\underline{n}}\end{array}
  :
  (m : \natkind) \to (\delta : \datatype) \to \tyexp[m.\mathsf{num}\langle \underline{n} \rangle] \to \tyexp[m\underline{n}.\delta]\quad (\text{where } \mathsf{num}\langle \underline{n} \rangle \text{ is a vector type})\mathrightfill
\end{displaymath}

\subsection{Translating Dot-product to OpenCL}
We pick up the dot product example~(\ref{code-ex:dot-product-complex}) given in \autoref{sec:motivation} to show how a mild variation which makes use of the OpenCL-specific primitives is translated to real OpenCL.
The example shown here uses the $\prim{mapWorkgroup}$ and $\prim{mapLocal}$ primitives together with the vectorisation primitives $\prim{asVector}$ and $\prim{asScalar}$.
\begin{displaymath}
\begin{array}[t]{@{}l}
  \prim{asScalar_4}~(\prim{join}~(\prim{mapWorkgroup}\\
  \qquad\qquad\qquad\qquad\begin{array}[t]{@{}l}
    (\lambda \mathit{zs}_1.~\prim{mapLocal}~(\lambda\mathit{zs}_2.~\prim{reduce}~
      (\lambda x~a.~(\prim{fst}~x * \prim{snd}~x) + a)~0~(\prim{split}~8192~\mathit{zs}_2))~\mathit{zs}_1) \\
     (\prim{split}~8192~(\prim{zip}~(\prim{asVector_4}~\mathit{xs})~(\prim{asVector_4}~\mathit{ys}))))))
  \end{array}
\end{array}
\end{displaymath}
This is the code used in the experimental evaluation (\autoref{sec:experimentalResults}) and shows excellent performance on an Intel CPUs compared to the reference MKL implementation. Vectorisation is crucial on Intel CPUs for achieving high performance.

This purely functional program with OpenCL-specific primitives is translated to the following imperative program.
The translation largely follows the steps explained in \autoref{sec:translation} extended to cover the OpenCL-specific primitives, as explained above.
\begin{displaymath}
    \begin{array}[t]{@{}l}
        \prim{parforWorkgroup}~(N / 8192)~(\prim{joinAcc}~(N/8192)~64~(\prim{asScalarAcc_4}~(N / 128)~\mathit{out}))~(\lambda~ gid~o .\\
        \enspace \prim{parforLocal}~64~o~(\lambda~lid~o .\\
        \enspace \enspace \prim{newPrivate}~\mathsf{num\langle 4 \rangle}~\mathit{accum}.\\
        \enspace \enspace \enspace \mathit{accum}.1 := 0;\\
        \enspace \enspace \enspace \prim{for}~2048~(\lambda~i.\\
        \enspace \enspace \enspace \enspace \mathit{accum}.1 :=
          \mathit{accum}.2~+\\
        \enspace \enspace \enspace \enspace \enspace ( \prim{fst}~(\prim{idx}~(\prim{idx}~(\prim{split}~2048~(\prim{idx}~(\prim{split}~(8192*4)~(\prim{zip}~(\prim{asVector_4}~xs)~(\prim{asVector_4}~ys)))~gid))~lid)~i) )~*\\
        \enspace \enspace \enspace \enspace \enspace ( \prim{snd}~(\prim{idx}~(\prim{idx}~(\prim{split}~2048~(\prim{idx}~(\prim{split}~(8192*4)~(\prim{zip}~(\prim{asVector_4}~xs)~(\prim{asVector_4}~ys)))~gid))~lid)~i) )\ );\\
        \enspace \enspace \enspace \mathit{out} := \mathit{accum.2}\ ) )\\
    \end{array}
\end{displaymath}
We generate the following OpenCL kernel where each line corresponds to a line of the imperative DPIA program.
\begin{lstlisting}[mathescape]
kernel void KERNEL(global float *out, const global float *restrict xs,
                   const global float *restrict ys, int N) {
 for (int g_id = get_group_id(0); g_id < N / 8192; g_id += get_num_groups(0)){$\label{line:parforWorkgroup}$
  for (int l_id = get_local_id(0); l_id < 64; l_id += get_local_size(0)){$\label{line:parforLocal}$
    float4 accum;
    accum = (float4)(0.0, 0.0, 0.0, 0.0);
    for (int i = 0; i < 2048; i += 1) {
      accum = (accum +
               (vload4(((2048 * l_id) + (8192 * 4 * g_id) + i), xs) *$\label{line:vload1}$
                vload4(((2048 * l_id) + (8192 * 4 * g_id) + i), ys))); }$\label{line:vload2}$
    vstore4(accum, ((64 * g_id) + l_id), out); } } }$\label{line:vstore}$
\end{lstlisting}
The $\prim{parforWorkgroup}$ and $\prim{parforLocal}$ primitives have been translated into $\prim{for}$ loops in line~\ref{line:parforWorkgroup} and~\ref{line:parforLocal} which use the OpenCL functions \texttt{get\_group\_id} and \texttt{get\_local\_id} for distributing iterations across parallel executing work-groups and work-items.
Loading elements as vector data types from the \texttt{float} arrays \texttt{xs} and \texttt{ys} requires using the OpenCL provided function \texttt{vload4} in lines~\ref{line:vload1} and~\ref{line:vload2}.
Similarly, storing the computed value with vector data type in the output array uses the \texttt{vstore4} function in line~\ref{line:vstore}.

\subsection{Memory allocation in Data Parallel Idealised Algol for OpenCL}
Our translation from functional to imperative programs leaves us with programs which perform statically bounded memory allocation.  The lifetime of every memory allocation is known because it is bounded by the scope of the $\prim{new}$ primitive.  Nevertheless, the memory allocation occurs dynamically as part of the execution of the program.  In C these allocations can be performed with \texttt{malloc} on the heap or \texttt{alloca} on the stack. However, OpenCL does not support dynamic memory allocation.  Furthermore, OpenCL demands that all temporary buffers in global and local memory -- even with statically known size -- have to be allocated prior to the kernel execution and passed as pointers to the kernel function.  In order to generate valid OpenCL, we perform an additional translation step to hoist all $\prim{newGlobal}$ and $\prim{newLocal}$ primitives to the very top of the program where we will eventually turn them into kernel arguments.  $\prim{new}$ primitives can be nested inside parallel for loops, so when hoisting memory allocations out of the loop the amount of memory has to be multiplied by the number of loop iterations, so that every loop iteration has its distinct location to write to.

To hoist the allocations we traverse the imperative program and for each parallel for loop we encounter we remember the number of iterations and the loop variable.  Once we reach a $\prim{newGlobal}$ or $\prim{newLocal}$ primitive, we replace it with its body and substitute the appropriate acceptor-expression pair for its variable that correctly points to the right place in the globally allocated data structure.

The following imperative \DPIA program implements dot-product with two memory allocations nested in the $\prim{parforGlobal}$ loop.
The allocation in global memory has to be hoisted out while the nested allocation in private memory ($\prim{newPrivate}$) is permitted in OpenCL, and will translate to the allocation of a scalar stack variable.
\begin{displaymath}
    \quad\begin{array}[t]{@{}l}
        \prim{parforGlobal}~n~\mathit{out}~(\lambda i~o .\\
        \quad \prim{newGlobal}~1024.\mathsf{num}~\mathit{tmp}.\\
        \quad \quad \prim{for}~1024~(\lambda j.~ \prim{idx}~\mathit{tmp}.1~j := (\prim{idx}~(\prim{idx}~(\prim{split}~1024~\mathit{xs})~i)~j) * (\prim{idx}~(\prim{idx}~(\prim{split}~1024~\mathit{ys})~i)~j)\ );\\
        \quad \quad \prim{newPrivate}~\mathsf{num}~\mathit{accum}.\\
        \quad \quad \quad \mathit{accum}.1 := 0;~\prim{for}~1024~(\lambda j.~\mathit{accum}.1 := \mathit{accum}.2 + (\prim{idx}~\mathit{tmp}.2~j)\ );~\mathit{out} := \mathit{accum.2}\ )\\
    \end{array}\mathrightfill
\end{displaymath}
To hoist out the allocation in global memory we first visit $\prim{parforGlobal}$, remember the number of iterations $n$ and the loop variable $i$.
Then, we replace the $\prim{newGlobal}$ with its body in which we have replaced $\mathit{tmp}$ with $(\prim{idx}~\mathit{tmp'}~i)$. We indicate the places where uses of $\mathit{tmp}$ have been replaced by shaded backgrounds:
\begin{displaymath}
    \quad\begin{array}[t]{@{}l}
        \prim{newGlobal}~(n\times 1024).\mathsf{num}~\shade{\mathit{tmp'}}.\\
        \quad \prim{parforGlobal}~n~\mathit{out}~(\lambda i~o .\\
        \quad \quad \prim{for}~1024~(\lambda j.~\prim{idx}~\shade{(\prim{idx}~\mathit{tmp'}.1~i)}~j := (\prim{idx}~(\prim{idx}~(\prim{split}~1024~\mathit{xs})~i)~j) * (\prim{idx}~(\prim{idx}~(\prim{split}~1024~\mathit{ys})~i)~j)\ );\\
        \quad \quad \prim{newPrivate}~\mathsf{num}~\mathit{accum}.\\
        \quad \quad \quad \mathit{accum}.1 := 0;~\prim{for}~1024~(\lambda j.~\mathit{accum}.1 := \mathit{accum}.2 + (\prim{idx}~\shade{(\prim{idx}~\mathit{tmp'}.2~i)}~j)\ );~\mathit{out} := \mathit{accum.2}\ )\\
    \end{array}\mathrightfill
\end{displaymath}
A $\prim{newGlobal}$ primitive is introduced at the very top of the program with the adjusted type.

\section{Experimental Results}
\label{sec:experimentalResults}

This section evaluates the quality of the OpenCL code generated from DPIA following the translation described in \autoref{sec:translation}.
We are interested to see if this formal translation introduces overheads compared to manual written OpenCL code and to the informal translations from functional programs to OpenCL used in \citep{SteuwerFeLiDu2015/icfp}, where semantic preserving rewrite rules were used purely at the functional level to explore different implementations.
We start by describing our experimental setup and the benchmarks used.


\newcommand\bench[1]{\emph{#1}}

\subsection{Experimental Setup}

With the help of the original authors, \citep{SteuwerFeLiDu2015/icfp}, we reproduced their results using the same methodology as them.
We used three three different OpenCL platforms:
1) an Nvidia GeForce GTX TITAN X with CUDA 8 and driver 375.26 installed;
2) an AMD Radeon HD 7970 GPU with AMD-APP 3.0 and driver 15.300 installed;
3) an Intel Xeon E5530 CPU with 8 physical cores distributed across two sockets and hyper-threading enabled.

We used the same set of benchmarks with two input sizes.
For \emph{scal}, \emph{asum}, and \emph{dot}, we used vectors of 16 (small) and 128 (large) millions elements.
For \emph{gemv}, input matrices of $4096^2$ (small) and $8192^2$ (large) elements were used.

We used the OpenCL profiling API for measuring OpenCL kernel runtime and the CPU runtime was measured using the \textit{gettimeofday} function.
We did not measure data transfer time, as we were only interested in the quality of the generated OpenCL kernel.
Each experiment was repeated 1000 times and we report median runtimes.
We compare against the manually written and optimised code from the vendor-provided libraries: CUBLAS version 8.0 from Nvidia, clBLAS version 2.12 from AMD, and MKL version 11.1 from Intel.

\begin{figure*}[t]
  \centering
  \begin{subfigure}[b]{0.315\linewidth}
    \includegraphics[width=\linewidth]{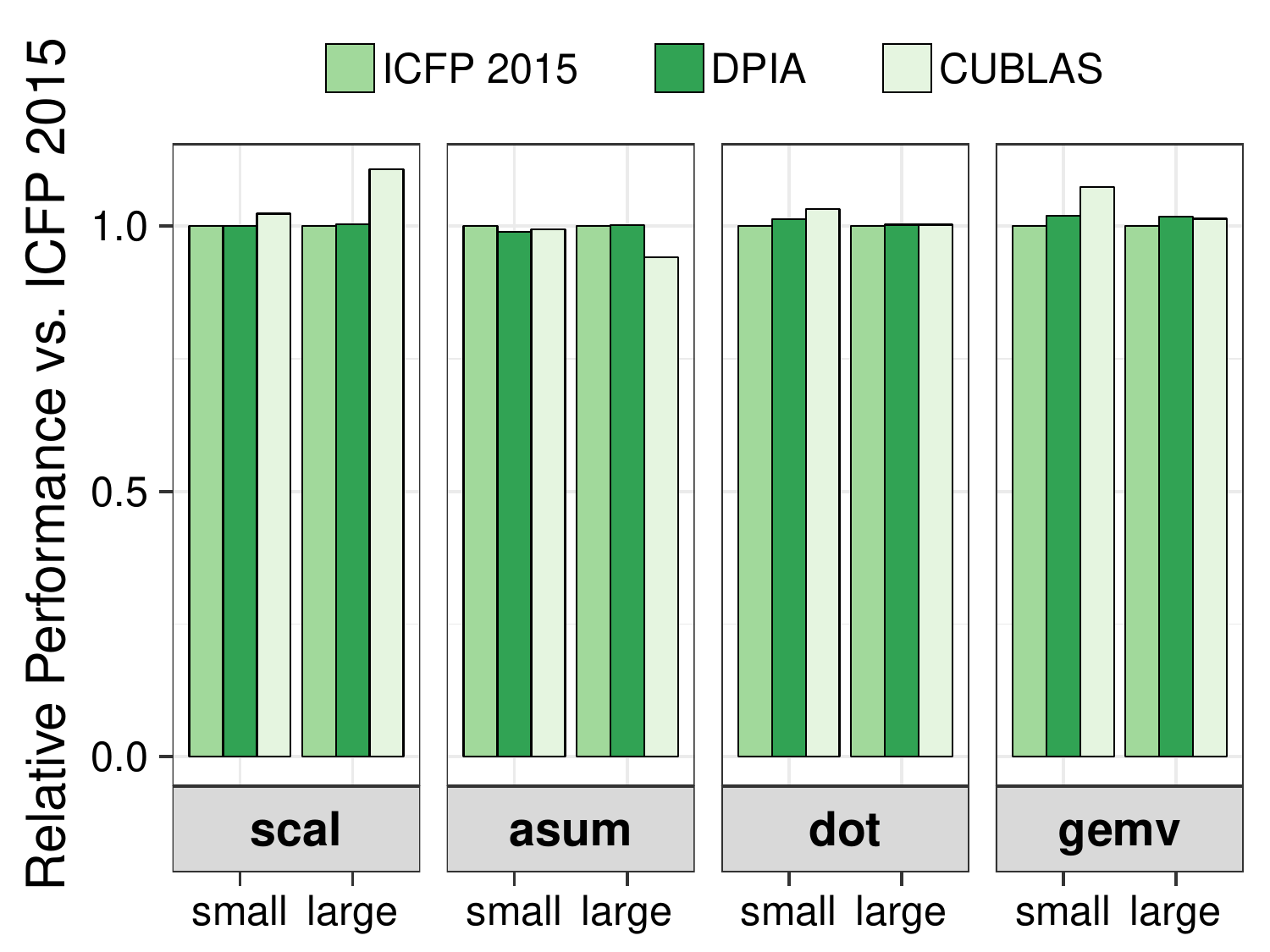}
    \caption{Nvidia GPU}
    \label{fig:results-nvidia}
  \end{subfigure}
  \hspace{0.015\linewidth}
  \begin{subfigure}[b]{0.315\linewidth}
    \includegraphics[width=\linewidth]{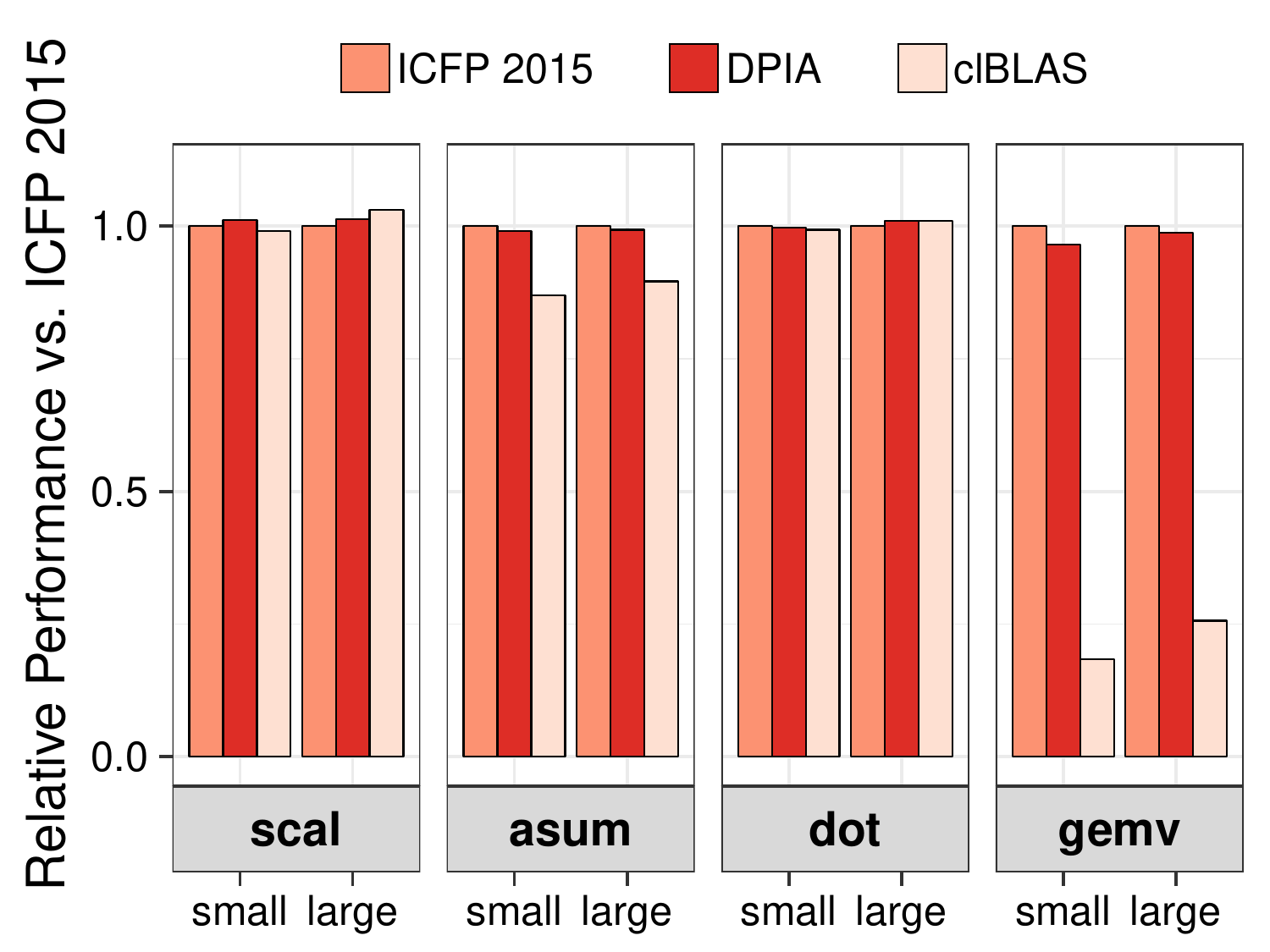}
    \caption{AMD GPU}
    \label{fig:results-amd}
  \end{subfigure}
  \hspace{0.015\linewidth}
  \begin{subfigure}[b]{0.315\linewidth}
    \includegraphics[width=\linewidth]{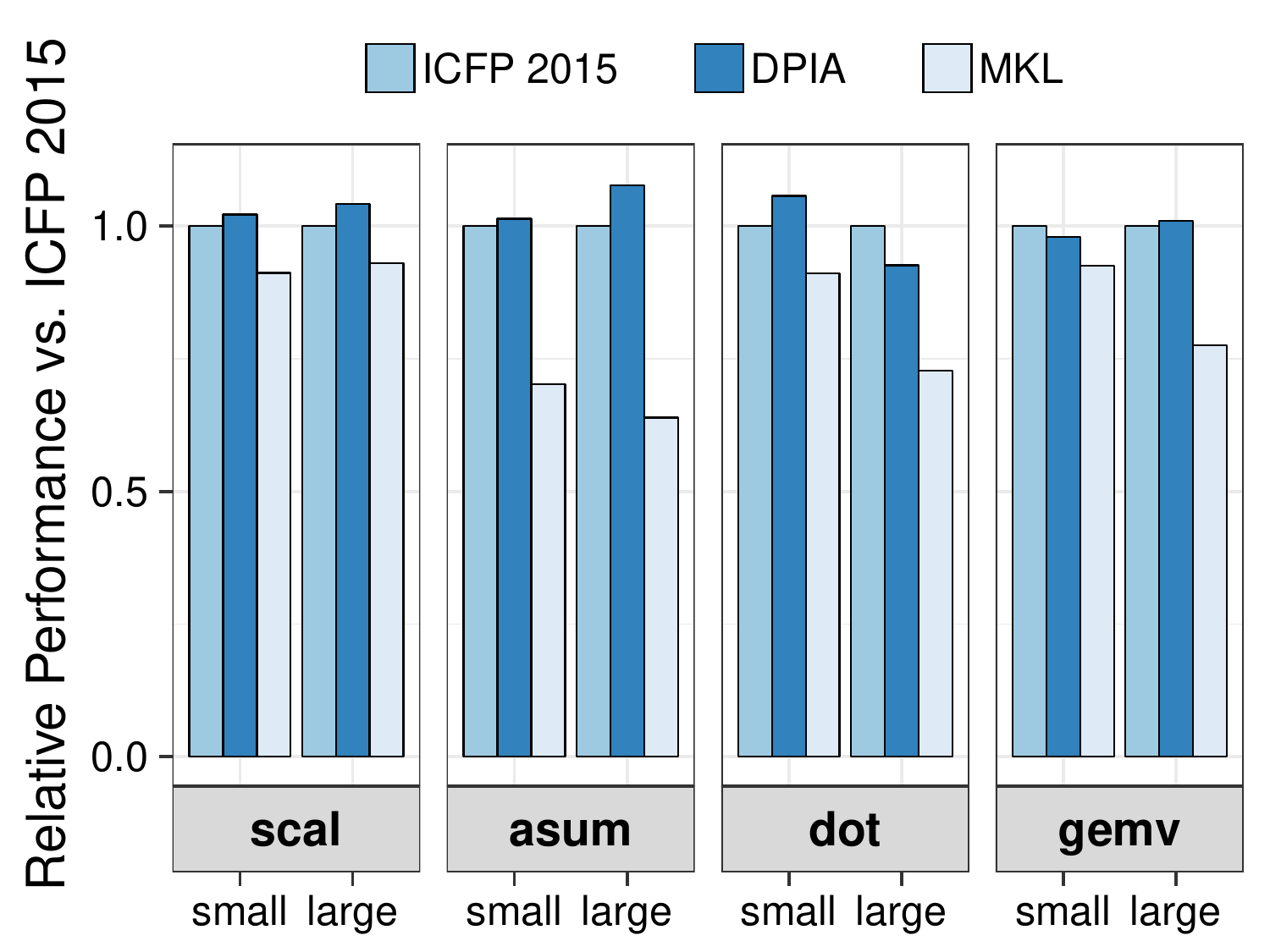}
    \caption{Intel CPU}
    \label{fig:results-intel}
  \end{subfigure}
  \caption{Performance comparison of code compiled via the formal translation from DPIA to OpenCL vs. informal translation of ICFP~2015~\citep[cf.][]{SteuwerFeLiDu2015/icfp} and vs. platform-specific libraries.
  The formal translation from DPIA to OpenCL introduces no performance overhead compared to ICFP~2015 and matches or outperforms highly tuned libraries on all three platforms.
         }
   \label{fig:results}
\end{figure*}

\subsection{Overhead of Formal Translation}
\autoref{fig:results} shows the runtime performance of the OpenCL kernels generated via the formal translation described in \autoref{sec:translation}.
The graphs are normalised by the performance of the OpenCL code generated from the technique described by \citet{SteuwerFeLiDu2015/icfp} (labelled ICFP 2015). Bars lower than 1.0 indicate a performance loss and bars higher than 1.0 a performance gain.

The performance of the OpenCL code generated by the method of \citeauthor{SteuwerFeLiDu2015/icfp} and the code generated from DPIA is almost identical in all cases with less than 5\% difference. This demonstrates that our formal translation process does not introduce significant overheads.

\subsection{Performance Comparison vs. Platform-Specific Libraries}
The performance results comparing DPIA generated OpenCL kernels against platform-specific libraries provided by Nvidia, AMD, and Intel show that for most benchmarks and input sizes the generated code matches the library performance.
For some cases, such as gemv on AMD or asum on Intel we even clearly outperform the library implementations by a factor of up to five times.
These performance results are similar to the results published by \citet{SteuwerFeLiDu2015/icfp} and show that by exploring parallelisation strategies using semantics preserving rewrite rules it is possible to outperform manually written code.
In this paper, we have extended the formal rewriting from the purely functional to the imperative level while achieving the same impressive performance results.



\section{Related Work}
\label{sec:relatedWork}

\paragraph{Idealised Algol and Syntactic Control of Interference}
We have heavily relied on Reynolds' insightful design of Idealised Algol (IA) in this work, originally spelt out in \cite{Reynolds97}. IA's orthogonal combination of typed $\lambda$-calculus and imperative programming has given us the ideal language in which to formalise compilation from functional to imperative code. Moreover, Reynolds' Syntactic Control of Interference (SCI) \cite{Reynolds78} enabled us to ensure that we always produce deterministic data race free programs. Brookes describes a concurrent version of Idealised Algol \cite{DBLP:journals/iandc/Brookes02} that he calls ``parallel'', but is intentionally a non-deterministic concurrent language that allows threads to communicate through shared memory.

Reynolds presents SCI as a series of principles for a language design, which are formulated as a substructural type system by \citet{OHearnPTT99}. We have used \citeauthor{OHearnPTT99}'s formulation in our design of \DPIA. We do not know of any other work using IA or SCI as an intermediate language for compilation, although Ghica has used interference controlled Idealised Algol as a high-level language for hardware synthesis~\citep{Ghica07}.

Reynolds' original presentation of IA describes its semantics in terms of a functor categories \cite{oles82category}. This semantics models the stack discipline of IA (which Reynolds uses to systematically derive a compilation strategy for IA~\citep{DBLP:conf/popl/Reynolds95}), but does not model non-interference or the locality of fresh state, both of which we rely upon in \autoref{sec:correctness}. O'Hearn and Tennent were the first to use relational parametricity to model locality \citep{DBLP:journals/jacm/OHearnT95}. Models of non-interference are given by \citet{DBLP:journals/mscs/OHearn93},\citet{DBLP:journals/iandc/Tennent90}, and, \citet{DBLP:journals/iandc/OHearnT93}, refining the original Reynolds/Oles functor category semantics. The semantic insights in this work later fed into Separation Logic \cite{DBLP:conf/popl/IshtiaqO01}. Reddy's object-space semantics~\citep{DBLP:journals/lisp/Reddy96,reddy94passivity} also modelled non-interference, but took an intensional viewpoint based on modelling interactions with the store, rather than transformations of the store. Reddy later synthesised the relationally parametric and intensional approaches in his automata-theoretic model~\citep{DBLP:journals/entcs/Reddy13}, which we used in \autoref{sec:correctness}.

\paragraph{Functional Compilation Approaches Targeting Heterogeneous Architectures}
There exist multiple functional approaches for generating code for heterogeneous hardware.
\citet{SteuwerFeLiDu2015/icfp} use a data parallel language similar to the functional subset of DPIA.
Semantics preserving rewrite rules are used to explore the space of possible implementations showing that achieving high performance across multiple architectures from functional code is possible.
Obsidian~\citep{SvenssonNS16} is a functional low-level GPU programming language which gives programmers flexibility over how to write efficient GPU code while still providing functional abstractions.
Accelerate is a Haskell embedded domain specific language providing higher level abstractions with the aim of generating efficient GPU code~\citep{ChakravartyKLMG11,mcdonell13optimising}.
LiquidMetal~\citep{dubach12lime} targets Field-Programmable Gate Array (FPGAs) and GPUs by extending Java with a data-flow programming model with built-in functional map and reduce operators.
\citet{BergstromR12} compile \textsc{Nesl}, which is a first-order dialect of ML supporting nested data-parallelism and introduced by \citet{blelloch93nessl}, to GPU code.
Nvidia implement NOVA~\citep{collins14nova}, a functional language targeted at code generation for GPUs, and Copperhead~\citep{catanzaro11copperhead}, a data parallel language for GPUs embedded in Python.
Delite~\citep{sujeeth14delite} is  a system that enables the creation of domain-specific languages using functional parallel patterns and targets multi-core CPUs and GPUs.

Our work is the first to formally explain the translation of a functional data parallel language to imperative code and to demonstrate that this does not introduce overhead compared to the existing state-of-the-art.

\section{Conclusions and Future Work}
\label{sec:conclusion}

This paper has introduced the new data parallel programming language \DPIA, an interference controlled dialect of Idealised Algol.
We showed how this language is used as a foundation for a formal translation of data parallel functional programs to imperative code for parallel machines.
This approach offers strong guarantees about the absence of data-races in the generated programs and offers a straightforward translation strategy for generating efficient OpenCL code.
Moreover, our approach is predictable and parallelism strategy preserving, allowing us to reuse \citet{SteuwerFeLiDu2015/icfp}'s automatic functional approach to generating high performance parallel strategies.
Our experimental results show that this formalised approach is able to produce high performance code on a par with existing ad hoc techniques without introducing any overheads.

Although we have captured aspects of the OpenCL parallelism and memory hierarchies in \DPIA (as described in \autoref{sec:opencl}, we are not guaranteed by construction to have generated legal programs that respect the hierarchy.
For example, nesting a $\prim{mapWorkgroup}$ inside a $\prim{mapLocal}$ should not be permitted.
In future work, we intend to capture this kind of ``hardware paradigm'' information in the type system of \DPIA.
The applicability of such a system goes beyond OpenCL.
Heterogeneous parallel architectures abound: from data centre sized clusters of machines to FPGAs.
A language based approach to describing such hierarchies will provide a framework for building robust strategy preserving functional to imperative compilation methods.
We also plan to push our formalisation future to prove correctness of the final translation from \DPIA to C and OpenCL.





\bibliography{references}



\end{document}